\documentclass[transmag]{IEEEtran}
\usepackage{latexsym}
\usepackage[utf8]{inputenc}
\usepackage{cite}
\usepackage{hyperref}
\usepackage{graphicx}
\usepackage{tikz}
	\usetikzlibrary{arrows.meta}
	\usetikzlibrary{shapes}
\usepackage{pgfplots}
	\usepgfplotslibrary{groupplots}
	\pgfplotsset{compat=newest,every axis/.append style={font=\footnotesize}}
\usepackage{amsfonts,amssymb,amsmath}
\usepackage{url}
\usepackage{algorithmic}
\usepackage{array}
\usepackage[caption=false,font=normalsize,labelfont=sf,textfont=sf]{subfig}
\usepackage{etoolbox}
\usepackage[normalem]{ulem}

    \def\natu{\mathbb{N}}
    \def\reals{\mathbb{R}}
    \def\integ{\mathbb{Z}}
    
    \def\rationals{\mathbb{Q}}
    
    \newcommand{\listintzero}[1]{\lbrace 0, 1, \dots, #1\rbrace}
	
    \newcommand{\mse}[1]{\operatorname{\mathrm{MSE}}( #1 )}
    \def\datarv{\mathbf{Y}}
    \def\datasa{\mathbf{y}}
    \def\model{\mathcal{Y}_{\pars}}
    \newcommand{\pdf}[3]{\operatorname{f_{#1}}\!\left(#3;#2\right)}
    
    \newcommand{\normal}[2]{\operatorname{\mathcal{N}}\!\left(#1,#2\right)}
    \newcommand{\uniform}[2]{\operatorname{\mathcal{U}}\!#1,#2}
    \newcommand{\mean}[1]{\boldsymbol{\mu}_{#1}}
    
    \def\pars{\boldsymbol{\theta}}
    \def\parsone{\pars^{(1)}}
    \def\parstwo{\pars^{(2)}}

    \newcommand{\trans}[1]{{#1}^\mathrm{T}}
    \newcommand{\Id}[1]{\mathbf{\mathrm{I}}_{#1}}
    \newcommand{\ones}[1]{\mathbf{1}_{#1}}
    
    \renewcommand{\mod}[1]{\operatorname{mod}_{#1}\!}
    \def\m2pi{\mod{2\pi}}
    \def\modu{\mod{1}}
    \newcommand{\extrema}[3]{\operatorname{#1}_{#2}\left\lbrace #3 \right\rbrace}
    
    \renewcommand{\max}[2]{\extrema{max}{#1}{#2}}
    \renewcommand{\min}[2]{\extrema{min}{#1}{#2}}
    
    \def\freqd{f_{\mathrm{d}}} 
    
    \def\master{\mathcal{M}} 
    \def\slave{\mathcal{S}}
    \def\samplingperiod{T_{\mathrm{s}}} 
    \def\slaveperiod{T_{\slave}} 
    \def\masterperiod{T_{\master}}
    \def\delay{\delta_0} 
    \def\delayping{\delta_{\rightarrow}}
    \def\delaypong{\delta_{\leftarrow}}
    \def\delaytrans{\delta_{\leftrightarrow}}
    \def\slavephase{\phi_{\slave}} 
    \def\masterphase{\phi_{\master}}
    \def\SNRin{\mathrm{SNR}_\mathrm{in}}
    \def\SNRout{\mathrm{SNR}_\mathrm{out}}

    \newtheorem{lemma}{Lem.}
    \newtheorem{theorem}{Theorem}
    \newtheorem{definition}{Def.}
	
    \def\figsnew{figs}
    
      \def\n{\mathbf{n}}

\begin{document}

  \title{Clock synchronization over networks --- Identifiability of the sawtooth model}
  \author{
    Pol~del~Aguila~Pla, \IEEEmembership{Member, IEEE}, 
    Lissy~Pellaco, \IEEEmembership{Student Member, IEEE},
    Satyam~Dwivedi, \IEEEmembership{Member, IEEE}\\ 
    Peter~Händel, \IEEEmembership{Senior Member, IEEE}, 
    and Joakim~Jaldén, \IEEEmembership{Senior Member, IEEE}
    \thanks{
    Manuscript received October 23, 2019; revised February 18, 2020; accepted
    February 29, 2020. 
    The associate 
    editor coordinating the review of this manuscript and approving it for publication
    was Prof. Christos Masouros. 
    \emph{(Corresponding author: Pol del Aguila Pla)}}
    \thanks{This work was supported by the SRA ICT TNG project Privacy-preserved Internet Traffic Analytics (PITA).}
    \thanks{Pol del Aguila Pla is with the Center for Biomedical Imaging, in Switzerland, and with 
    the Biomedical Imaging Group, École polytechnique fédérale de Lausanne, Switzerland (email: pol.delaguilapla@epfl.ch).
    Pol performed the work while at the KTH Royal Institute of Technology.}
    \thanks{
    Lissy Pellaco, Peter Händel and Joakim Jald\'{e}n are with the Division of Information
    Science and Engineering, School of Electrical Engineering and Computer Science, KTH Royal Institute of 
    Technology, Stockholm, Sweden (e-mail: pellaco@kth.se, ph@kth.se, and jalden@kth.se). }\thanks{Satyam Dwivedi is 
    with Ericsson Research, Stockholm, Sweden (e-mail: dwivedi@kth.se).}%
    \thanks{This paper has supplementary downloadable material available at http://ieeexplore.ieee.org., provided by the authors. The material includes derivations that, although lengthy or challenging, have not been included in the paper. This material is 245 KB in size.}
    }

\IEEEtitleabstractindextext{\begin{abstract} In this paper, we analyze the two-node joint clock 
synchronization and ranging problem. We focus on the
case of nodes that employ time-to-digital converters
to determine the range between them precisely. This
specific design choice leads to a sawtooth model 
for the captured signal, which has not been studied
before from an estimation theoretic standpoint. In the
study of this model, we recover the basic conclusion
of a well-known article by Freris, Graham, and Kumar
in clock synchronization. More importantly, we 
discover a surprising identifiability result on the 
sawtooth signal model: noise improves the theoretical
condition of the estimation of the phase and offset
parameters. 
To complete our study, we provide performance references
for joint clock synchronization and ranging using the
sawtooth signal model by presenting an exhaustive 
simulation study on basic estimation strategies under 
different realistic conditions. 
With our contributions in this paper, we enable 
further research in the estimation of sawtooth signal models
and pave the path towards their industrial use for 
clock synchronization and ranging.
\end{abstract}%
\begin{IEEEkeywords}%
Clock synchronization, ranging, identifiability, sawtooth model, sensor networks, round-trip time (RTT).
\end{IEEEkeywords}%
}

\newcommand{\new}[1]{{#1}}

\newcommand{\old}[1]{}

\maketitle

\section{Introduction}

    \IEEEPARstart{C}{lock} synchronization across a deployed
  network is a pervasive and long-standing 
  challenge~\cite{Lamport1978,Freris2010,Etzlinger2014,Xia2018,Geng2018,Etzlinger2018}. 
  Furthermore, new-generation technologies each require more
  accurate synchronization. To name a few, i) in cellular
  communications, synchronization between base stations through a backhaul channel is
  fundamental to maintain frame alignment and permit handover
  among neighboring cells, and has been identified as a crucial
  requirement for distributed beamforming, interference 
  alignment, and user 
  positioning~\cite{Zachariah2017,Koivisto2017}, 
  ii) in radio-imaging technology~\cite{Event2019}, accurate
  clock synchronization between the sparse chips that form an
  array is critical, and, in active-sensing $3$-dimensional
  cases~\cite{Chen2017,Jamali2018}, it results in low-cost 
  wide-aperture ultra-short ultra-wideband (UWB) pulses,
  increasing both the angular and depth resolutions of the
  captured images, 
  iii) in wireless sensor networks~\cite{He2014,Carli2014},
  synchronization is critical to data-fusion, channel-sharing,
  coordinated scheduling~\cite{Kim2017,Bolognani2016}, and
  distributed control~\cite{Freris2010} and iv) in distributed
  database solutions that provide external consistency, clock
  synchronization accuracy regulates latency, throughput, and
  performance~\cite{Geng2018}. 
  
  Consequently with this wide range of application, theoretical
  insights on the fundamental limitations of clock
  synchronization over networks are likely to incite radical
  innovations in a number of fields. In~\cite{Freris2011},
  Freris, Graham, and Kumar established the fundamental
  limitations of the clock synchronization problem in an
  idealized scenario. Particularly, given a network of nodes with
  noise-less affine clocks and fixed unknown link delays that
  exchange time-stamped messages, \cite{Freris2011} 
  i)~showed that clock synchronization was only possible if the 
  link delays were known to be symmetric, and
  ii)~characterized the uncertainty regions of the clock
  synchronization parameters under different hypotheses.
  In this paper, we analyze the same problem from a perspective
  that is closer to real implementation. In short, we analyze the
  two-node joint clock synchronization and ranging 
  problem~\cite{Zheng2010,Chepuri2013,Dwivedi2015} with noisy
  round-trip time (RTT) measurements without 
  time-stamps~\cite{Gholami2015}, for a node design originally
  proposed in~\cite{Angelis2013} to improve ranging accuracy. 
  The resulting analysis has several advantages. 
  First, because protocols without time stamps require only minimal 
  transmissions of very short pulses carrying no information, the resulting
  technology minimizes communication overhead, and is beneficial in applications in which the data-rates
  are critically needed for other uses~\cite[p.~29]{Etzlinger2018}. 
  Second, because we 
  consider hardware specifically tailored to ranging accuracy, 
  we reveal how applications that require this accuracy, such as 
  cooperative localization~\cite{Patwari2005}, 
  positioning~\cite{Nilsson2014}, and control~\cite{Freris2010},
  can harness the same hardware and protocols for
  synchronization. 
  Third, the analysis is more realistic, because it takes into account 
  the real-world stochasticity of the measurements. In 
  particular, in our analysis of the
  problem we i) unveil the need for symmetric delays in RTT-based
  protocols, in a direct parallel to the discovery 
  in~\cite{Freris2011}, 
  ii) find novel results on the identifiability of sawtooth
  signal models under diverse conditions, which are of interest
  by their own right to chaotic system 
  analysis~\cite{Ciftci2001,Drake2007} and control, and 
  iii) provide performance references to guide practitioners in
  their use of this technology.
  
  In summary, in this paper we first derive from basic principles
  a model for RTT measurements between two nodes equipped with
  time-to-digital converters (TDC) in a network with fixed,
  unknown link delays (Theorem~\ref{th:deterministic_model}). 
  Then, we shift our focus towards an encompassing family of
  signal models, i.e., sawtooth signal models, when one considers 
  different stochastic effects. In this context, we provide 
  results on the identifiability of these models, both negative 
  (Lemma~\ref{lem:unid-nojit}) and positive 
  (Theorem~\ref{th:id}), under different noise conditions. 
  Here, we obtain the surprising result that the presence of a
  noise term inside a non-linear model term makes said model
  identifiable.
  We then shift the focus again towards
  clock synchronization and ranging, 
  and we provide a thorough and verifiable empirical evaluation 
  of the basic estimation techniques we propose in~\cite{AguilaPla2020}
  to exploit the sawtooth model 
  (Figs.~\ref{fig:vs_freq_d_randomized}--\ref{fig:vs_SNRs_randomized}, 
  implementation accessible in \cite{GITHUB}). 
  These empirical results, together with the approximated Cram\'{e}r-Rao
  lower bounds derived in~\cite{AguilaPla2020}, are clear and simple 
  performance references for clock synchronization and ranging using sawtooth
  models. Such performance references are of use to both engineers
  that use this technology and to researchers aiming to develop estimation
  techniques for sawtooth signal models.
  
\subsection{Notation}
  \label{subsec:Notation}
  
  Discrete random processes will be in uppercase letters and square brackets, such as $Y[n]$, while deterministic sequences, 
  e.g., realizations of said processes, will be lowercase with square brackets, i.e., $y[n]$. For both these sequences, 
  the notation will be simplified by omitting the discrete time index when it can be established by context. Vector random 
  variables will be bold uppercase letters, e.g., $\datarv$, while deterministic vectors
  will be bold lowercase letters, e.g., $\datasa \in\reals^N$. 
  Functions, on the other hand, will be non-italics lower-case letters, e.g., the probability 
  density function (PDF) of a vector random variable $\datarv$ on a parametric family with vector parameter $\pars$ will be 
  $\pdf{\datarv}{\pars}{\datasa}$. Through the paper, we view the modulus equivalence as a function, i.e., we note the 
  mapping $x\mapsto y \in [0,a) \mid (y = x) \mod{}\, (a) $ as 
  $\mod{a}:\reals \rightarrow [0,a)$.

\section{The sawtooth model} \label{sec:SysMod}

  In applications in which high ranging accuracy with low communication overhead is desired, a low-cost solution 
  in terms of both complexity and power consumption is using node designs that include TDCs to measure RTTs~\cite{Angelis2013}. 
  Indeed, such sensors were successfully incorporated in a prototype system to aid firefighters by providing on-site 
  infrastructure-free indoor positioning~\cite{Nilsson2014}.
  TDCs, however, induce an asymmetry between the rate at which nodes can measure time and the rate at which they can 
  act upon their environment. This asymmetry generates an unexpected waveform in the sequence of RTT measurements over time. 
  This phenomenon was first reported by~\cite{Dwivedi2015}, where a sawtooth model was proposed and empirically validated, 
  and possible applications to clock synchronization over networks were identified.
  In this section, we reintroduce the design of~\cite{Angelis2013} and derive the sawtooth model from a few simple assumptions.
  
  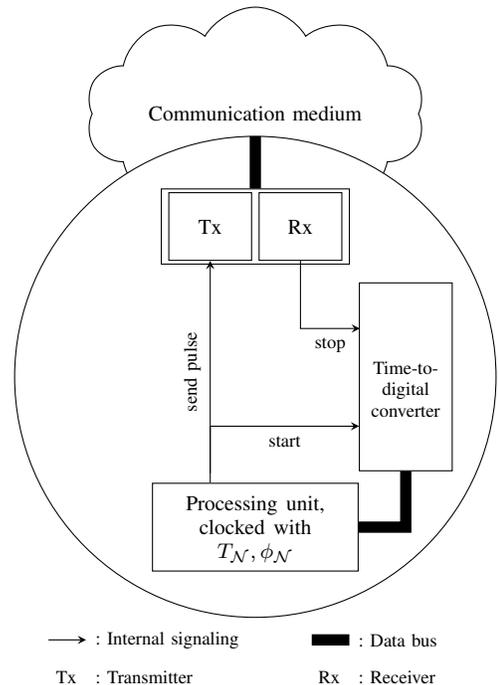
\begin{figure}
	  \centering
	  \begin{tikzpicture}
  \node (cloud) at (0,3.5) [cloud, draw,cloud puffs=10,cloud puff arc=120, aspect=2, inner ysep=1em] 
    {\footnotesize Communication medium};
  \draw [fill=white] (0,0) circle (3.2cm);
  \node (PU) [draw, rectangle, minimum width=2.5cm, minimum height=1cm, text width=2.5cm] 
  at (0,-2) {\vspace{-8pt}\begin{center} \footnotesize Processing unit, clocked with 
  $T_{\mathcal{N}},\phi_{\mathcal{N}}$\end{center}};
  \node (TRX) [draw, rectangle, minimum width=2.5cm, minimum height=1cm]
  at (0,2) {};
    \node (TX) [draw, rectangle, minimum width=1.1cm, minimum height=0.9cm]
    at (-0.6,2) {\footnotesize Tx};
    \node (RX) [draw, rectangle, minimum width=1.1cm, minimum height=0.9cm]
    at (0.6,2) {\footnotesize Rx};
  \draw [line width=1.5mm] (TRX.north) -- (0,3.2);
  \node (TDC) [draw, rectangle, minimum width=0.8cm, minimum height=2.5cm, text width=1cm]
  at (2,0) {\begin{center} \scriptsize Time-to-digital converter \end{center}};
  \draw [stealth-] (TX.south) --++ (0,-2.95) node (PUstart){} 
  node [midway,above,rotate=90] {\scriptsize send pulse};
  \draw [line width=1.5mm] (PU.east) -| (TDC.south);
  \draw (PUstart) --++ (0,.75) node (startbreak){};
  \draw [-stealth] (startbreak.center) --++ (1.98,0) node [midway,below] {\scriptsize start};
  \draw [-stealth] (RX.south) --++ (0,-.9) --++ (.79,0) node [midway,below] {\scriptsize stop};
    \node[anchor=west] (Signal) at (-2.25,-3.5) {\scriptsize: Internal signaling};
    \draw [stealth-] (Signal.west) --++ (-0.5,0);
    \node[anchor=west] (Bus) at (1.25,-3.5) {\scriptsize: Data bus};
    \draw [line width=1.5mm] (Bus.west) --++ (-0.5,0);
     \node[anchor=west] (RXlab) at (1.25,-4) {\scriptsize: Receiver};
    \node[anchor=east] at (RXlab.west) {\scriptsize Rx};
    \node[anchor=west] (TXlab) at (-2.25,-4) {\scriptsize: Transmitter};
    \node[anchor=east] at (TXlab.west) {\scriptsize Tx};
\end{tikzpicture}
	  
	  \vspace{-5pt}
	  
	  \caption{
	    Internal design of any node $\mathcal{N}$ considered throughout this paper.
	    Initially proposed in \cite{Angelis2013}, this design uses an independent time-to-digital
	    converter (TDC) to accurately measure round-trip times (RTT). The resulting node 
	    $\mathcal{N}$ can measure RTTs with a much finer time-resolution than it can react to 
	    incoming pulses. Indeed, in order to react to an incoming pulse, $\mathcal{N}$ must first 
	    access the TDC's memory, process the reading, and decide to send a pulse, all of which 
	    require waiting until its next clock cycle. \label{fig:node_internals}
	  }
  \end{figure}
  Consider now the design of \cite{Angelis2013}, described in Fig.~\ref{fig:node_internals}.
  Here, each node or sensor $\mathcal{N}$ has a processing unit, a transceiver and a TDC. 
  With this design, a sensor can measure RTTs at the resolution of the TDC, usually in 
  the order of $\mathrm{ps}$, much finer than the period of the processing unit's clock, usually
  in the order of tens of $\mathrm{ns}$. Besides the clear advantage of this design for 
  ranging through RTT measurements, this creates an interesting asymmetric behavior of the node
  as an agent and as a measuring device. As we will show below (Theorem~\ref{th:deterministic_model}), this asymmetry produces a sawtooth
  waveform in the measured RTTs that depends on the synchronization parameters, and, under
  reasonable assumptions, leads to a viable system for clock synchronization over networks. As a consequence, such a design may also be considered for wired networks, where ranging information is usually not relevant. A final by-product of the inclusion of a TDC in the design in Fig.~\ref{fig:node_internals} is that we can consider that each node has perfect knowledge of its own clock period, which 
  it measures directly with its TDC.
  
  \subsection{Deterministic model} \label{sec:ideal_model}

  Consider two nodes, $\master$ and $\slave$, designed as $\mathcal{N}$ in Fig.~\ref{fig:node_internals}, in a network (wired or wireless).
  These two nodes execute the RTT measurement scheme illustrated in Fig.~\ref{fig:RTT}. In this scheme, 
  $\master$ measures the RTT between itself and $\slave$ by sending pulses (a.k.a. pings) to $\slave$ and 
  using its TDC to accurately record when a response (a.k.a. pong) is received from $\slave$. In particular,
  $\master$ sends a pulse at some of the times at which its clock has upflanks, i.e., at the times 
  $t_n = K \masterperiod n$. Here $\masterperiod>0~[\mathrm{s}]$ is 
  $\master$'s clock period, the sampling factor $K\in\natu$ is designed to determine the sampling period 
  $\samplingperiod = K \masterperiod~[\mathrm{s}]$, $n\in\natu$ is a discrete-time index, and we assume 
  without loss of generality that $\master$'s clock phase offset is zero, i.e., 
  $\masterphase=0~[\mathrm{rad}]$. 
  \begin{figure}
    \centering
      \begin{tikzpicture}[xscale=1.1]
	\node[draw,circle,minimum size=20pt,color=black,fill=white,inner sep=0pt] (mas) at (0,0) {\footnotesize $\master$};
	\node at (0,-0.55) {\footnotesize $ \masterperiod, \phi_{\master}$};
	\node[draw,circle,minimum size=20pt,color=black,fill=white,inner sep=0pt] (sla) at (4,0) {$\slave$};
	\node at (4,-0.55) {\footnotesize $ \slaveperiod, \phi_{\slave}$};
	\draw[thick,orange,dotted,-Stealth,opacity=0.7] (1,0.1) -- (3,0.1) node [midway,above]{\footnotesize ping};
	\draw[thick,blue,loosely dashed,Stealth-,opacity=0.7] (1,-0.1) -- (3,-0.1) node [midway,below]{\footnotesize pong};
	\draw[dotted,thin] (mas.north) -- (0,1.2);
	\draw[dotted,thin] (sla.north) -- (4,1.2);
	\draw[dashdotted,<-] (0,.65) -- (4,.65) node [midway,above] {\footnotesize $\delaypong$};
	\draw[dashdotted,->] (0,1.15) -- (4,1.15) node [midway,above] {\footnotesize $\delayping$};
\end{tikzpicture}
      \def\xs{.97} \def\ys{.8}

\begin{tikzpicture}[xscale=\xs,yscale=\ys]
      \def\zes{3}
      \def\ons{4}
      \def\Ps{0.8}
      \def\NP{8}
      \node [anchor=east] at (0,.5*\zes+.5*\ons) {\footnotesize $\mathcal{S}$'s clock};
      \draw [gray,thick,-Latex] (0,\zes) -- (0,\ons+.5);
      \draw [gray,thick,-Latex] (0,\zes) -- (\NP*\Ps+\Ps+.2,\zes) node [anchor=north] {\footnotesize $t$};
      \foreach \p in {0,1,...,\NP}{
	\draw [thin] (\p*\Ps,\zes) -- (\p*\Ps,\ons) -- (\p*\Ps+.5*\Ps,\ons) -- (\p*\Ps+.5*\Ps,\zes) -- (\p*\Ps+\Ps,\zes);
      }
      \draw [<->] (3*\Ps,\ons+.1) -- (4*\Ps,\ons+.1) node [midway, above] {\footnotesize $\slaveperiod$};
      \draw[densely dotted] (3*\Ps,\ons+.2) -- (3*\Ps,\ons); \draw[densely dotted] (4*\Ps,\ons+.2) -- (4*\Ps,\ons);
      \def\zem{0}
      \def\onm{1}
      \def\Pm{0.75}
      \def\NP{8}
      \def\Ts{2}
      \def\transf{1.5}
      \def\d0{\Ps}
      \node [anchor=east] at (0,.5*\zem+.5*\onm) {\footnotesize $\mathcal{M}$'s clock};
      \draw [gray,thick,-Latex] (0,\zem) -- (0,\onm+.5);
      \draw [gray,thick,-Latex] (0,\zem) -- (\NP*\Pm+\Pm+.2,\zem) node [anchor=north] {\footnotesize $t$};
    \foreach \p in {0,1,...,\NP}{
      \draw [thin] (\p*\Pm,\zem) -- (\p*\Pm,\onm) -- (\p*\Pm+.5*\Pm,\onm) -- (\p*\Pm+.5*\Pm,\zem) -- (\p*\Pm+\Pm,\zem);
    }
    \def\p{9}
    \draw [thin,densely dotted] (\p*\Pm,\zem) -- (\p*\Pm,\onm) -- (\p*\Pm+.5*\Pm,\onm) -- (\p*\Pm+.5*\Pm,\zem) -- (\p*\Pm+\Pm,\zem);
    \draw [<->] (3*\Pm,\zem-.1) -- (4*\Pm,\zem-.1) node [midway, below] {\footnotesize $\masterperiod$};
    \draw[densely dotted] (3*\Pm,\zem-.2) -- (3*\Pm,\zem); \draw[densely dotted] (4*\Pm,\zem-.2) -- (4*\Pm,\zem);
    \node at (-0.5,2) {\footnotesize messages};
    \foreach \p in {0,1,2}{
      \draw [orange,dotted,-Stealth,opacity=0.7] (\p*\Pm*\Ts,\onm) -- (\p*\Pm*\Ts+\transf,\zes) node [pos=0.15,above] {\tiny ping};
      \node [orange,opacity=0.7] at (\p*\Pm*\Ts+\transf,\zes) {\tiny $ \bullet $};
    }
    \draw [orange,<->,thick,opacity=0.7] (0,\zem-.1) -- (\Pm*\Ts,\zem-.1) node [midway,below] {\footnotesize $\samplingperiod$};
    \draw [orange, densely dotted] (0,\zem) -- (0,\zem -.2); \draw[orange, densely dotted] (\Pm*\Ts,\zem) -- (\Pm*\Ts,\zem-.2);
     
    \draw [blue,loosely dashed,-Stealth,opacity=0.7] (2*\Ps+\d0,\zes) -- (2*\Ps+\transf+\d0,\onm) node [pos=0.4,above] {\tiny pong};
      \node [blue,opacity=0.7] at (2*\Ps+\transf+\d0,\onm) {\tiny $\bullet$};
    \draw [blue,loosely dashed,-Stealth,opacity=0.7] (4*\Ps+\d0,\zes) -- (4*\Ps+\transf+\d0,\onm) node [pos=0.4,above] {\tiny pong};
      \node [blue,opacity=0.7] at (4*\Ps+\transf+\d0,\onm) {\tiny $\bullet$};
    \draw [blue,loosely dashed,-Stealth,opacity=0.7] (6*\Ps+\d0,\zes) -- (6*\Ps+\transf+\d0,\onm) node [pos=0.4,above] {\tiny pong};
      \node [blue,opacity=0.7] at (6*\Ps+\transf+\d0,\onm) {\tiny $\bullet$};
    
    \draw [blue,<->,thick] (2*\Ps,\zes) -- (2*\Ps+\d0,\zes) node [midway,below] {\footnotesize $\delay$};
    \draw [blue,<->,thick] (4*\Ps,\zes) -- (4*\Ps+\d0,\zes) node [midway,below] {\footnotesize $\delay$};
    \draw [blue,<->,thick] (6*\Ps,\zes) -- (6*\Ps+\d0,\zes) node [midway,below] {\footnotesize $\delay$};
    
    \foreach \p/\c in {0/magenta,1/green!50!black,2/red!50!black}{
       
      \draw [|<->|,thick,\c] (\p*\Pm*\Ts,\onm-0.1-.4*\p) -- (2*\p*\Ps+2*\Ps+\transf+\d0,\onm-0.1-.4*\p) node [pos=0.25,below,yshift=2pt] {\footnotesize $y[\p]$};
    }
\end{tikzpicture}
      
      \vspace{-18pt}
	  
    \caption{ Example of the round-trip time (RTT) measurement scheme in Section~\ref{sec:ideal_model}.
    $\master$ sends ping pulses to $\slave$ at its clock upflanks
    every $\samplingperiod = 2\masterperiod$. The ping is recorded in $\slave$'s time-to-digital converter
    (TDC). At its next clock upflank,
    $\slave$ accesses its TDC and starts a delay of $\delay = \slaveperiod$ before responding with a pong pulse.
    The pong is recorded in $\master$'s TDC as soon as it arrives. \label{fig:RTT} }
  \end{figure}
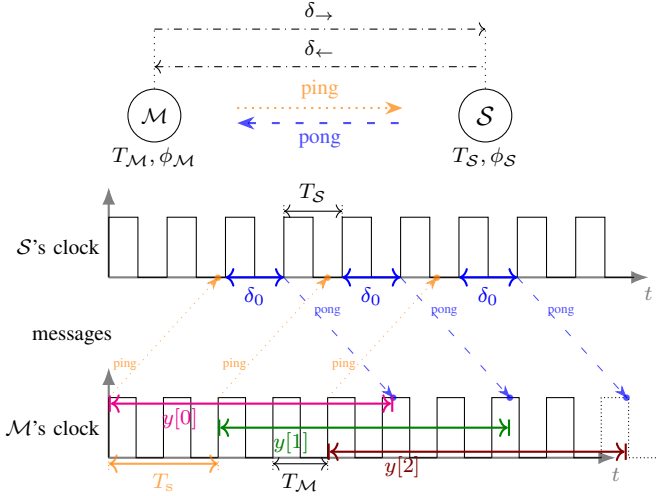
  \begin{figure*}
    \centering
    \input{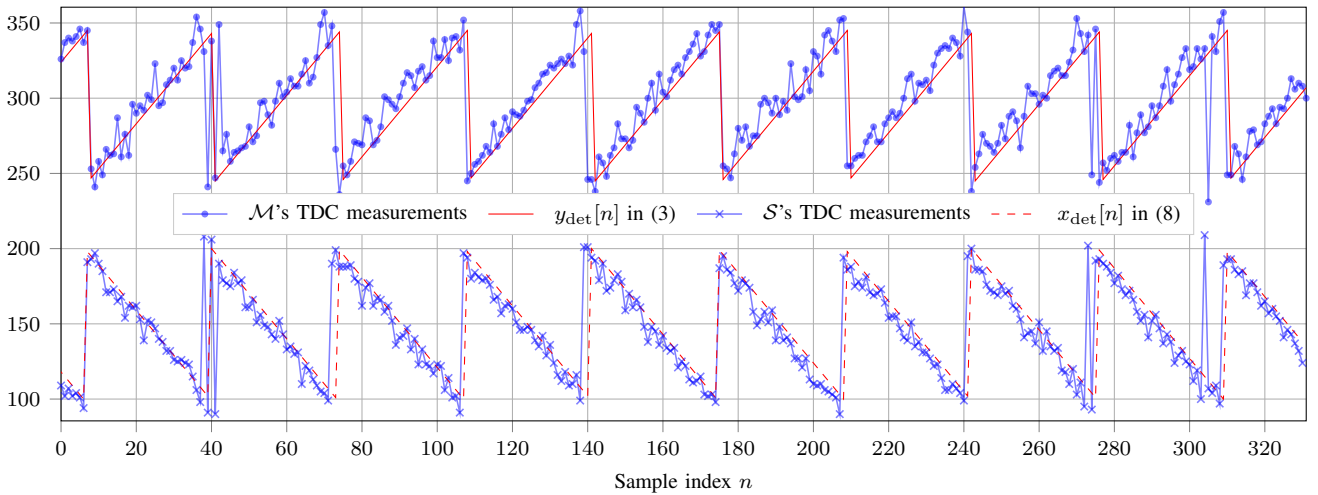}
    
    \vspace{-10pt}
    
    \caption{ TDC measurement buffer, in units of the TDC's clock, taken by
			$\master$ and $\slave$ in a simulation of the protocol described by 
			Fig.~\ref{fig:RTT} and Section~\ref{sec:ideal_model}. Unrealistic parameters
			were used to obtain a cheap-to-compute, simple, representative figure. 
      For more details on how this simulation was performed see this project's 
      repository at~\cite{GITHUB}. \label{fig:M_and_S} }
  \end{figure*}
  If we assume that the delays involved in the pulse traveling from $\master$ to $\slave$ accumulate to a constant value
  $\delayping~[\mathrm{s}]$, the $n$-th pulse arrives at $\slave$ and is recorded in its TDC at time $t_n+\delayping$.
  Nonetheless, $\slave$ will not be able to access the TDC's memory before its next clock upflank, and consequently, any action by $\slave$ will
  be further delayed until $t_n + \delayping + \Delta_n$. Here, $\Delta_n\geq 0~[\mathrm{s}]$ is the time remaining until $\slave$'s next 
  clock upflank. If we consider that $\slave$ has a clock with period $\slaveperiod>0~[\mathrm{s}]$ and phase $\slavephase~[\mathrm{rad}]$, i.e., an 
  offset delay of $\varphi_{\slave} = \slaveperiod \slavephase / (2\pi)~[\mathrm{s}]$, then $\slave$ has its clock upflanks at those times 
  that are at an integer number of periods away from $\varphi_\slave$, i.e., at the times
  $\tau\geq0$ when $\mod{\slaveperiod}\left( \tau + \varphi_\slave \right) = 0$.
  Furthermore, we know that $\Delta_n \leq \slaveperiod$, as $\slaveperiod$
  is the time between consecutive upflanks. Consequently, to obtain a closed-form expression for $\Delta_n$, we need to find 
  \begin{IEEEeqnarray*}{c}
	  \Delta_n = \min{}{ 
		  \tau \in(0,\slaveperiod) : 
		  \mod{\slaveperiod}\left( t_n + \delayping + \tau + \varphi_\slave \right) =0 }\mbox{.}
  \end{IEEEeqnarray*}
  Because $\mod{a}\left(b+c\right) = \mod{a}\left( \mod{a}[b] + \mod{a}[c]\right)$ for any $a\geq 0$ and $b,c\in\reals$, the condition for $t_n + \delayping + \tau$ to be the time of one of $\slave$'s clock 
  upflanks can be rewritten as $\tau = Q\slaveperiod - \mod{\slaveperiod}\left( t_n+\delayping +\varphi_\slave \right)$ for some 
  $Q\in\integ$. Then, because $\mod{\slaveperiod}\left( t_n+\delayping +\varphi_\slave \right)< \slaveperiod$ and $\tau \in (0,\slaveperiod)$, we conclude that
  \begin{IEEEeqnarray}{c} \label{eq:Delta_n}
	  \Delta_n = \slaveperiod - \mod{\slaveperiod}\left( t_n+\delayping +\varphi_\slave \right)\,.
  \end{IEEEeqnarray}
  We now allow for a known delay $\delay~[\mathrm{s}]$ to be introduced by $\slave$, which can account for any processing required
  to read the TDC's state and prepare the new pulse, and will usually be an integer number of $\slave$'s clock periods, i.e., 
  $\delay = K_0 \slaveperiod$. 
  Finally, as we did for the ping pulse, we consider $\delaypong$ to express the fixed delay for a pong pulse from $\slave$
  to reach $\master$ and be captured by the TDC. In conclusion, if we disregard the effect of the resolution of the TDC, which is usually
  four orders of magnitude finer than that of the nodes' clocks, the $n$-th RTT measurement will amount to
  \begin{IEEEeqnarray}{c} \label{eq:model_discussion_end}
	  y_{\mathrm{det}}[n] = \delay + \delaytrans + \Delta_n \,,
  \end{IEEEeqnarray}
  where $\delaytrans = \delayping + \delaypong$. We summarize our result in the following theorem.

  \begin{theorem}[Deterministic RTT measurement model] \label{th:deterministic_model}
	  Consider two nodes $\master$ and $\slave$ designed as specified in Fig.~\ref{fig:node_internals}.
	  Then, if $\master$ and $\slave$ follow the RTT measurement protocol specified above, 
	  $\freqd = 1/\slaveperiod - 1/\masterperiod$ and $\delaytrans = \delayping+\delaypong$, the $n$-th RTT
	  measurement $y_{\mathrm{det}}[n]$ can be expressed as
	  \begin{IEEEeqnarray}{rl} \label{eq:ideal_model}
		  y_{\mathrm{det}}[n] \,& = \delaytrans + \delay + \slaveperiod h[n] \mbox{, where } \\
		  h[n] \,& = 1 - \mod{1}\left( \samplingperiod \freqd n + \frac{\delayping}{\slaveperiod} + \frac{\slavephase}{2\pi} \right)\,. \nonumber
	  \end{IEEEeqnarray}
  \end{theorem}
  \begin{IEEEproof}
	  From \eqref{eq:Delta_n} we have that
	  \begin{IEEEeqnarray}{rl}
		  \Delta_n \,& = \slaveperiod - \mod{\slaveperiod}\left( \samplingperiod n+\delayping +\varphi_\slave \right)  \\
			  \label{eq:proof_ideal_model:samplingperiod}
			  & = \slaveperiod \left(1 - \mod{1}\left(K \frac{\masterperiod}{\slaveperiod} n 
			    + \frac{\delayping}{\slaveperiod} + \frac{\slavephase}{2\pi}  \right)\right) \\
			  \label{eq:proof_ideal_model:periodicity}
			  & = \slaveperiod \left(1 - \mod{1}\left( K \frac{\masterperiod - \slaveperiod}{\slaveperiod} n 
			    + \frac{\delayping}{\slaveperiod} + \frac{\slavephase}{2\pi}  \right) \right) \\
			  \label{eq:proof_ideal_model:fddef}
			  & = \slaveperiod \left(1 - \mod{1}\left(\samplingperiod \freqd n 
			    + \frac{\delayping}{\slaveperiod} + \frac{\slavephase}{2\pi}  \right)\right)\,.
	  \end{IEEEeqnarray}
	  Here, we have used that $\samplingperiod = K \masterperiod$ and $\varphi_\slave = \slaveperiod \slavephase / (2\pi)$ in 
	  \eqref{eq:proof_ideal_model:samplingperiod}, that $\mod{1}$ is periodic with period one in \eqref{eq:proof_ideal_model:periodicity},
	  and that $\freqd=1/\slaveperiod - 1/\masterperiod$ in \eqref{eq:proof_ideal_model:fddef}. Finally, \eqref{eq:ideal_model} follows 
	  from substituting \eqref{eq:proof_ideal_model:fddef} in
	  \eqref{eq:model_discussion_end}.
	  
  \end{IEEEproof}
  In conclusion, by running the RTT protocol specified in Fig.~\ref{fig:RTT}, $\master$ obtains 
  data intimately related with the parameters it needs to predict $\slave$'s clock signal, i.e., to synchronize to $\slave$. Indeed, having measured its own clock period $\masterperiod$ using its TDC, $\freqd$ reveals $\slaveperiod$, which together with $\slavephase$ characterizes $\slave$'s clock signal completely. The goal of our study is to establish under which conditions $\master$ will be able to simultaneously estimate these parameters.
  
  Other formulations of the model \eqref{eq:ideal_model} in terms of the usual synchronization parameters for affine clocks, i.e. the clock
  skew $\alpha_\slave = \slaveperiod / \masterperiod$ and the offset delay $\varphi_\slave$, including the general expression for when
  $\varphi_\master \neq 0$, can be found in the supplementary material to this paper. Nonetheless, the expression in \eqref{eq:ideal_model} remains 
  the most practical, because it expresses the compromise between the sampling period $\samplingperiod$ and the frequency difference $\freqd$ 
  of the system, which will prove to be relevant to our analysis.
  
  Incidentally, under the simple assumption that $\samplingperiod\geq\delta_0 + \slaveperiod$, which can be guaranteed under any reasonable 
  $\freqd$ if $K>K_0+1$, if we assume that $\slave$'s TDC starts measuring every time $\slave$ sends a pong and stops measuring when the next
  ping is received, the $n$-th measurement $x[n]$ taken by $\slave$'s TDC can be expressed as
  \begin{IEEEeqnarray}{c} \label{eq:model_s}
    x_{\mathrm{det}}[n] = \samplingperiod - \delay - \slaveperiod h[n]\,.
  \end{IEEEeqnarray}
  As we will see, this will imply that even while $\master$ is leading the RTT measurement protocol, $\slave$ could still perform frequency 
  synchronization. Nonetheless, we will not consider $\slave$'s TDC measurements for most of the paper, and we will instead focus on determining the conditions
  under which $\master$ can achieve full synchronization and ranging. More details on the derivations of~\eqref{eq:ideal_model}~and~\eqref{eq:model_s}
  can be found in the supplementary material to this paper.
  
  In this project's repository, accessible at ~\cite{GITHUB}, we validate \eqref{eq:ideal_model} by simulating an ideal physical system as 
  described above and verifying the exact correspondence between the model and the obtained measurements. In Fig.~\ref{fig:M_and_S}, we
  show the fits of \eqref{eq:ideal_model} and \eqref{eq:model_s} on the TDC measurements of $\master$ and $\slave$ throughout a simulated run
  with noisy clock periods and noisy transmission delays.
  
  \subsection{Stochastic model}
  
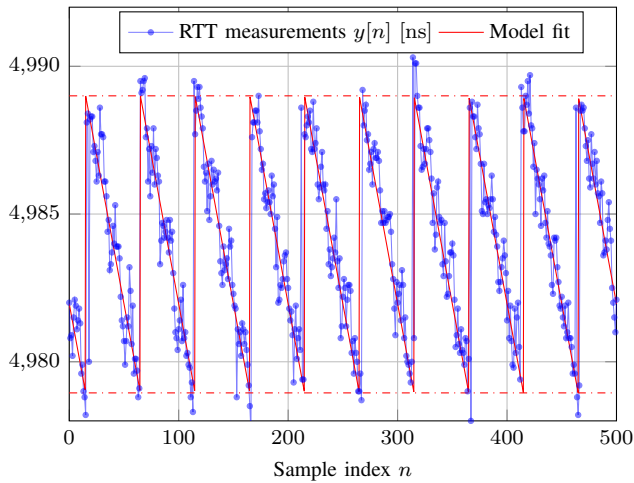
\begin{figure}
    \centering
    \begin{tikzpicture}
    \begin{axis}[
      height=0.8\columnwidth,
      width=\columnwidth,
      baseline,
      xlabel={Sample index $n$},
      grid=both, 
      xmin=0,
      xmax=500,
      ymin=4978,
      ymax=4992,
      legend cell align = {left},
      legend columns=-1,
      legend entries = {{RTT measurements $y[n]~[\mathrm{ns}]$},{Model fit}},
      legend style = {at={(0.94,0.99)}},
      ]
      \addplot[blue,mark=*,mark size=1pt,mark options={opacity=0.5},opacity=0.5,x filter/.code={\pgfmathparse{\pgfmathresult-1}}] table {\figsnew/sawtooth.dta};
      \addplot [samples=1000,domain=0:500,red,opacity = 1] {4989-1.6*mod(2*pi*x/50+4.4,2*pi)};
      \addplot[samples=2,domain=0:500,mark=none, red, dashdotted] {4989};
      \addplot[samples=2,domain=0:500,mark=none, red, dashdotted] {4989-1.6*2*pi};
    \end{axis}
\end{tikzpicture}
  
    \vspace{-10pt}

    \caption{RTT measurements from the ultra-wide band testbed from \cite{Angelis2013}, compared to a fit of the
    deterministic model \eqref{eq:ideal_model}. In dash-dotted horizontal lines, the maximum and minimum values 
    of the model. \label{fig:sawtooth}}
  \end{figure}

\begin{figure}
    \centering
    \input{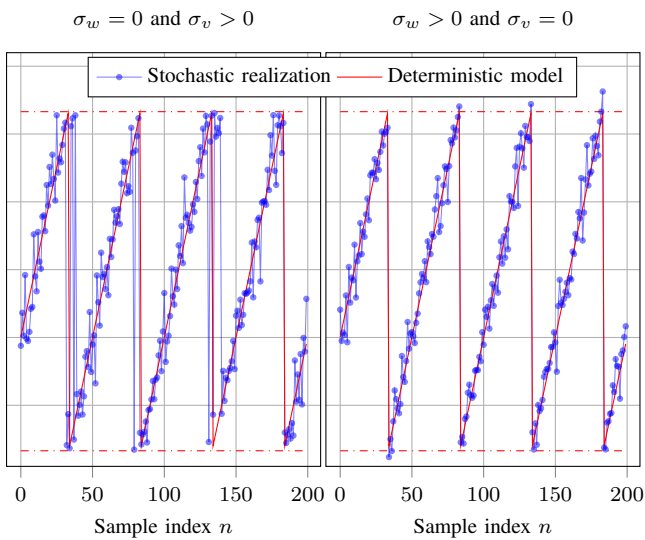}
    
    \vspace{-10pt}        
    
    \caption{ Two realizations of our stochastic model \eqref{eq:stoch_model},
      exemplifying the effects of additive noise inside and outside $\mod{1}(\cdot)$. 
      On one hand, $\sigma_w=0$ and $\sigma_v>0$ leads to many high jumps around the
      wrapping points. On the other hand, $\sigma_w>0$ and $\sigma_v=0$ leads to a 
      signal that is not bounded by the minimum and maximum values of the deterministic
      model \eqref{eq:ideal_model} (shown in dash-dotted horizontal lines). 
      For further examples of the effects of noise in measurements following our stochastic
      model \eqref{eq:stoch_model}, as well as the effects of randomness in the physical 
      quantities described above,
      see this project's GitHub repository at \cite{GITHUB}. 
    \label{fig:example_signals} }
  \end{figure}

  In Fig.~\ref{fig:sawtooth}, we show real RTT data obtained in \cite{Dwivedi2015} from the ultra-wide band testbed of~\cite{Angelis2013} 
  using this RTT scheme, accompanied by an example model fit.
  Given the observed signal and its expected shape, a simple observation is that $\slave$'s clock was faster than that of $\master$ in the specific experimental 
  set-up, because the ramps in the sawtooth signal have negative slope, which implies that $\freqd>0$.
  Fig.~\ref{fig:sawtooth} also exemplifies the two distinct effects that random deviations of the physical parameters
  can produce on the data. 
  On one hand, large jumps of approximately $\slaveperiod$ in the measured RTT are observed (effect i])
  if a random deviation influences the specific clock period at which $\slave$
  reads the arrival of a ping pulse from its TDC, i.e., it changes which is the first up-flank in $\slave$'s clock after the ping 
  pulse arrives. On the other hand, if this does not happen, random deviations appear directly in the signal as additive noise 
  (effect ii]). From a modeling perspective, these 
  two effects are not easily represented distinctively. Indeed, variations of the transmission time from $\master$ to $\slave$,
  $\delayping$, or jitter in any of the two clock periods, $\masterperiod$ or $\slaveperiod$, could lead to any of the two described effects, while variations
  of the transmission time from $\slave$ to $\master$, $\delaypong$, can only ever lead to effect ii]. In this paper, we will consider 
  the effect of random variations on the physical parameters, as well as the quantization by the TDC, in the form of two additive 
  white noise processes$W[n]$ and $V[n]$, inside and outside the 
  nonlinearity, respectively. 
  In short, our stochastic model for the RTT measurements taken by $\master$ is 
  \begin{IEEEeqnarray}{rl} \label{eq:stoch_model}
    Y[n] \,& = \delaytrans + \delay + W[n] + \slaveperiod H[n] \mbox{, where } \\
    H[n] \,& = 1 - \mod{1}\left( \samplingperiod \freqd n + \frac{\delayping}{\slaveperiod} + \frac{\slavephase}{2\pi} + V[n] \right)\,. \nonumber
  \end{IEEEeqnarray}
  For simplicity, we will assume that $W[n]$ and $V[n]$ are zero-mean Gaussian processes with respective standard deviations $\sigma_w$ and $\sigma_v$ and we will consider 
  them independent. Analyzing the effect of the existing dependence between them, or evaluating the magnitude or effect
  of this dependence, is outside of the scope of this paper. 
  Only adding $V[n]$, i.e., setting $\sigma_w=0$, could explain 
  the two effects explained above for most sample indices $n$.
  Nonetheless, as shown by the indicators of the maximum and minimum of the 
  model fit in Fig.~\ref{fig:sawtooth}, the experimental RTT measurements are
  not bounded in the range $[\delaytrans + \delay, \delaytrans + \delay + \slaveperiod]$, indicating that the noise term outside the non-linearity $W[n]$ is necessary. 
  Furthermore, random variations in $\delaypong$ or $\delay$ cannot be meaningfully represented by $V[n]$, since variations of these parameters
  of any magnitude will never affect which upflank of $\slave$ detects the ping pulse.
  Fig.~\ref{fig:example_signals} exemplifies the effect of each of the noise terms
  by showing two realizations of our stochastic model \eqref{eq:stoch_model}, one
  in which $\sigma_w=0$ and $\sigma_v>0$, and one in which $\sigma_w>0$ and 
  $\sigma_v=0$.
  
  In order to simplify the notation for the rest of the paper and abstract some of our theoretical results,
  we will express the stochastic sawtooth model in terms of four generic parameters,
  an offset $\alpha\in\reals$, a non-zero amplitude $\psi\in\reals\setminus\lbrace0\rbrace$ with known sign 
  $\mathrm{sign}(\psi)$, a normalized frequency $\beta \in [-1/2,1/2)$, and 
  a normalized phase offset $\gamma\in[0,1)$. In other words, we will express the sawtooth signal model as
  \begin{IEEEeqnarray}{rl}\label{eq:model}
    Y[n] \,& = \alpha + W[n] + \psi \modu\left(\beta n + \gamma + V[n] \right)\,,
  \end{IEEEeqnarray}
  with $W[n]$ and $V[n]$ independent additive white Gaussian noise processes.
  An empirical analysis verifying this model \eqref{eq:model} on real data from the testbed of~\cite{Angelis2013} 
  can be found in~\cite{Dwivedi2015}. Here, the restriction of the $\beta$ and $\gamma$ parameters simply reflects the 
  maximum ranges that we can expect to distinguish, given the periodicity of $\mod{1}\left(\cdot\right)$ as a function. Indeed, adding any 
  integer factor of $n$ inside the modulus, or any integer by itself, will not change $Y[n]$, and establishes an equivalence
  of period one for both $\beta$ and $\gamma$. Here, we have chosen $\beta\in[-1/2,1/2)$ and $\gamma\in[0,1)$ to preserve their 
  intuitive meanings as a normalized frequency and a phase term, respectively. Several initial insights can be drawn from the parallel between 
  \eqref{eq:stoch_model} and \eqref{eq:model}. First, the condition $\left| \freqd \right| < 1/(2\samplingperiod)$, resembling the Nyquist 
  sampling condition, arises from the restriction in $\beta$. Second, if we consider this restriction and examine the relation between the 
  parameters of both models, we observe that
  \begin{IEEEeqnarray*}{rl}
    \alpha\, &= \delay + \delaytrans + \slaveperiod\mbox{, }  \psi = - \slaveperiod\mbox{, }\beta = \freqd \samplingperiod\mbox{, and }\\
    \gamma\, &= \mod{1}\left( \frac{\delayping}{\slaveperiod} + \frac{\slavephase}{2\pi} \right)\,,
  \end{IEEEeqnarray*}
  and, incorporating that $\slaveperiod = \masterperiod / (\masterperiod \freqd + 1)$ and that 
  $\masterperiod$, $\samplingperiod$ and $\delay$ are known,
  \begin{IEEEeqnarray}{rl} \label{eq:conversion}
    \delaytrans \, &= \alpha - \delay - \frac{\masterperiod}{\masterperiod \freqd + 1} \mbox{, }\\
    \freqd \,& = \frac{\beta}{\samplingperiod} = -\left( \frac{1}{\masterperiod} + \frac{1}{\psi} \right)\mbox{, and }\\
    \slavephase \, &= 2\pi \mod{1}\left( \gamma - \mod{1}\left(  \delayping \frac{\masterperiod \freqd + 1}{\masterperiod} \right) \right)\,.
  \end{IEEEeqnarray}
  Clearly, then, unless further constraints relating
  $\delayping$, $\delaypong$ and $\freqd$ are given, it is impossible to 
  recover $\delayping$, $\delaypong$, $\freqd$, and $\slavephase$ from 
  $\alpha$, $\psi$, $\beta$ and $\gamma$. 
  In the context of clock synchronization over networks, this is 
  equivalent to the impossibility result of \cite{Freris2011}, which
  studied the uncertainty sets where the synchronization parameters are known to lie given time-stamped 
  message exchanges under different conditions. An analysis similar to that in \cite{Freris2011} under 
  idealized, noise-free conditions could be reproduced for \eqref{eq:ideal_model}, but is outside of the 
  scope of this paper. 
  In contrast, we will provide an analysis of identifiability when every physical parameter can be subject
  to noise. In fact, this analysis will reveal that synchronization with the sawtooth signal model
  requires a certain level of randomness, i.e., it is impossible without it. Consequently with the discussion above,
  then, we will assume that $\delayping$ is given when one knows 
  $\delaytrans$ and $\freqd$, as it happens in a number of applications. For example, in wireless sensor networks,
  one may generally consider that all nodes are equal and the channels between any two of them are 
  symmetric, and thereby one can assume $\delayping = \delaypong = \delaytrans/2 = \delta_1 + \rho/c$
  where $\delta_1>0~[\mathrm{s}]$ is a known delay, $\rho>0~[\mathrm{m}]$ is the unknown 
  range between $\master$ and $\slave$ in the communication medium and $c~[\mathrm{m/s}]$ is the speed of light 
  in the medium. Even in this context, true line-of-sight communication is not a requirement, and one only needs to assume that the multipath is not dense and the direct path is not fully blocked, as in such a case the TDC will trigger on the first pulse, corresponding to the shortest path.
  These assumptions are consistent with ultra-wide band pulses such as the ones used in \cite{Dwivedi2015}.
  When convenient in the paper, we will use this assumption combined with $\delta_1 \approx 0$, and consider 
  the ranging problem of \cite{Angelis2013,Nilsson2014} jointly with clock synchronization \cite{Dwivedi2015}.
  In the following section, we will characterize the model~\eqref{eq:model} statistically, providing conditions 
  for its identifiability. Our aims in doing that are 1) to present novel results on the sawtooth signal model, and 2) to 
  provide guarantees for the design of practical synchronization systems using nodes
  modeled by the design in Fig.~\ref{fig:node_internals}.

\section{Identifiability of the sawtooth model} \label{sec:identifiability}

    Identifiability is a basic requirement on any statistical model that relates to the
  minimal conditions that make parameter estimation a reasonable goal \cite{Walter1997}.
  \begin{definition}[Identifiability] \label{def:id}
    (From \cite[Definition 11.2.2, p. 523]{Casella2002})
    Let $\model$ be a statistical model with parameter 
    $\pars\in\Omega$. Assume that if 
    $\datarv\sim \model$
    for some given $\pars$, $\datarv$ has PDF $\pdf{\datarv}{\pars}{\datasa}$.
    Then, $\model$ is an identifiable model, and 
    $\pars$ is an identifiable parameter, if and only if
    \begin{IEEEeqnarray}{c} \label{eq:def:id}
      \pdf{\datarv}{\parsone}{\datasa}
      = \pdf{\datarv}{\parstwo}{\datasa},
      \forall \datasa \Leftrightarrow \parsone = \parstwo\,.
    \end{IEEEeqnarray}
    That is, the mapping between the parameter $\pars$ and the distribution 
    specified by $\model$ is one-to-one.
    
  \end{definition}
  If \eqref{eq:def:id} is not met, the data observed when the 
  parameter value is $\parsone$ and the data observed when the parameter value
  is $\parstwo$ have the same distribution, and therefore, distinction between these
  two parameters from observed data is impossible. Unintuitively, even if \eqref{eq:def:id}
  is not given, one could possibly design \emph{good} estimators for $\pars$. Specifically, as long as the selected metric 
  in the space of parameters $\Omega$ does not assign much importance to the difference between the pairs
  $\parsone$ and $\parstwo$ that do not fulfill \eqref{eq:def:id}, estimation could remain a sensible objective.
  In this paper, the data model $\model$ is defined by \eqref{eq:model},
  and the considered parameters are $\pars = \trans{\left[ \alpha, \psi, \beta, \gamma \right]}$. 
  Hence, this section will be dedicated to establishing under which conditions, 
  in terms of the values of $\sigma_w$ and $\sigma_v$ in \eqref{eq:model}, a 
  one-to-one relation between $\pars$ and the distribution of the data 
  $\datarv=\trans{[Y[0], Y[1], \dots, Y[N-1]]}$ can be ensured.

  \subsection{Unidentifiability without inner noise} \label{sec:id_nojit}

  In order to analyze the relation between $\pars$ and $\pdf{\datarv}{\pars}{\datasa}$,
  we will first consider the simplifying assumption $\sigma_v=0$. This is an unrealistic
  assumption under most applicable uses of the sawtooth model \eqref{eq:model}, including
  that of clock synchronization, but it will be useful for our analysis. We will show 
  that under this assumption, \eqref{eq:model} yields an unidentifiable model $\model$ in
  which the effect of $\alpha$ and $\gamma$ cannot be fully distinguished in the observed
  data $\datarv \sim \model$.
  
  \begin{figure}
    \centering
      \def\xs{0.96} \def\ys{.5} \def\fs{\footnotesize}

\def\pini{0.7}
\def\pshift{0.5*pi+0.2}
\def\xfinal{2*pi+1}
\pgfmathsetmacro\bythree{1/3}
\pgfmathsetmacro\byxsc{1/\xs}
\begin{tikzpicture}[xscale=\xs,yscale=\ys]
    \node at (-0.5,7.5) {\fs (a)};
    \node[anchor=west] (RhoName) at (pi*0.25,7.5) {\fs: $\Delta\alpha$};
    \draw [latex-,color=red, very thick, dotted] (RhoName.west) --++ (-0.5,0);
    \node[anchor=west,color=red] (RhoSign) at (RhoName.east) {\tiny $\bullet$};
    \node[anchor=west] at (RhoSign) {\fs: $\mean{\Delta\alpha}$ };
    \node[anchor=west] (PhaseName) at (pi*1.25,7.5) {\fs: $\Delta\gamma$};
    \draw [stealth-,color=olive, very thick, dashed] (PhaseName.west) --++ (-0.5,0);
    \node[anchor=west,color=olive] (PhaseSign) at (PhaseName.east) {\tiny $\blacksquare$};
    \node[anchor=west] at (PhaseSign) {\fs: $\mean{\Delta\gamma}$ };
        \draw [dotted,xstep=.5*pi,ystep=.5*pi, very thin] (0,0) grid (\xfinal,7);
        \draw [dashed,xstep=pi,ystep=pi, gray, very thin] (0,0) grid (\xfinal,7);
        \node at (\pini,0) [anchor=north] {\fs  $\gamma$};
        \draw (\pini,-0.1) -- (\pini,0);
        \draw [dotted,very thin] (\pini,0) -- (\pini,\pini);
        \node (EndX) at (\xfinal,0) [anchor=west] {};
        \node at (pi,-1) [anchor=north] {\fs $\beta n + \gamma$};
        \node (EndY) at (0,8) [anchor=south] {\fs $[\mean{\pars}]_n$};
        \draw[->] (-0.1,0) -- (EndX);
        \draw[->] (0,-0.2) -- (EndY);
        \foreach \x/\xl in {.5*pi/\tfrac{1}{4}, pi/\tfrac{1}{2},   0.5*3*pi/\tfrac{3}{4}, 2*pi/1 } 
        {
	        \draw (\x,0) node[anchor=north] {\fs $\xl$};
	        \draw (\x,-0.1) -- (\x,0);
	    }
        \foreach \y/\yl in {0/\alpha, .5*pi/\alpha+\tfrac{1}{4}\psi, pi/\alpha+\tfrac{1}{2}\psi,   0.5*3*pi/\alpha+\tfrac{3}{4}\psi, 2*pi/\alpha+\psi }
        {
            \draw (0,\y) node[anchor=east] {\fs $\yl$};
            \draw (-.1*.5*\byxsc,\y) -- (0,\y);
        }
        \draw[color=blue, thick, opacity=0.3] (0,0) -- (2*pi,2*pi) -- (2*pi,0) -- (2*pi+0.1, 0.1);
        \draw[color=blue, dashed, thick, opacity=0.3] (2*pi+0.1, 0.1) -- (\xfinal,\xfinal-2*pi);
        \foreach \p/\n in { \pini/0, 2*pi*\bythree+\pini/1}
        {
            \draw[-stealth,color=olive, very thick, dashed] (\p, \p) -- (\p+\pshift,\p+\pshift);
            \draw[-latex,color=red, very thick, dotted] (\p,\p) -- (\p,\p+\pshift);
            \draw[dashed,color=olive] (0,\p+\pshift) -- (\p+\pshift,\p+\pshift);
            \draw[dotted,color=red,thick] (0,\p+\pshift) -- (\p,\p+\pshift);
            \node at (0,\p+\pshift) [color=olive] {\tiny $\blacksquare$};
            \node at (0,\p+\pshift) [color=red] {\tiny $\bullet$};
            \node (point) at (\p,\p) {$\bullet$};
            \node at (point.south) [anchor=north west, yshift=10pt] {\footnotesize $n=\n$};
        }
        \def\p{4*pi*\bythree+\pini}
        \draw[-latex,color=red, very thick, dotted] (\p,\p) -- (\p,\p+\pshift);
        \draw[dotted,color=red,thick] (0,\p+\pshift) -- (\p,\p+\pshift);
        \draw[color=olive, very thick, dashed] (\p,\p) -- (2*pi,2*pi);
        \draw[color=olive, very thick, dashed, -stealth] (2*pi,0) -- (\p+\pshift,\p+\pshift-2*pi);
        \draw[dashed,color=olive] (0,\p+\pshift-2*pi) -- (\p+\pshift,\p+\pshift-2*pi);
        \node at (0,\p+\pshift-2*pi) [color=olive] {\tiny $\blacksquare$};
        \node at (0,\p+\pshift) [color=red] {\tiny $\bullet$};
        \node (point) at (\p,\p) {$\bullet$};
            \node at (point.south) [anchor=north west, yshift=10pt] {\footnotesize $n=2$};
        \draw[dashdotted] (4*pi*\bythree+\pini,0) -- (4*pi*\bythree+\pini,4*pi*\bythree+\pini);
        \draw[dashed,<->, very thick] (4*pi*\bythree+\pini,pi) -- (2*pi,pi);
        \node at (.5*4*pi*\bythree+.5*\pini+pi,pi) [anchor=north] {\small  $\epsilon_+$};
        \draw[dashed,<->,very thick] (pi,4*pi*\bythree+\pini) -- (pi,2*pi);
        \draw[dashdotted] (0,4*pi*\bythree+\pini) -- (4*pi*\bythree+\pini,4*pi*\bythree+\pini);
        \node at (pi,.5*4*pi*\bythree+.5*\pini+pi) [anchor=east] {\small $\psi \epsilon_+$};
\end{tikzpicture}
      
      \vspace{-10pt}
      
      \def\xs{0.96} \def\ys{.5} \def\fs{\footnotesize}
	    
\def\pini{0.7}
\def\pshift{0.7}
\def\xfinal{2*pi+1}
\pgfmathsetmacro\bythree{1/3}
\pgfmathsetmacro\byxsc{1/\xs}
\begin{tikzpicture}[xscale=\xs,yscale=\ys]
  \node at (-0.5,7.5) {\fs (b)};
    \node[anchor=west] (RhoName) at (pi*0.25,7.5) {\fs: $\Delta\alpha$};
    \draw [latex-,color=red, very thick, dotted] (RhoName.west) --++ (-0.5,0);
    \node[anchor=west,color=red] (RhoSign) at (RhoName.east) {\tiny $\bullet$};
    \node[anchor=west] at (RhoSign) {\fs: $\mean{\Delta\alpha}$ };
    \node[anchor=west] (PhaseName) at (pi*1.25,7.5) {\fs: $\Delta\gamma$};
    \draw [stealth-,color=olive, very thick, dashed] (PhaseName.west) --++ (-0.5,0);
    \node[anchor=west,color=olive] (PhaseSign) at (PhaseName.east) {\tiny $\blacksquare$};
    \node[anchor=west] at (PhaseSign) {\fs: $\mean{\Delta\gamma}$ };
        \draw [dotted,xstep=.5*pi,ystep=.5*pi, very thin] (0,0) grid (\xfinal,7);
        \draw [dashed,xstep=pi,ystep=pi, gray, very thin] (0,0) grid (\xfinal,7);
        \node at (\pini,0) [anchor=north] {\fs $\gamma$};
        \draw (\pini,-0.1) -- (\pini,0);
        \draw [dotted,very thin] (\pini,0) -- (\pini,\pini);
        \node (EndX) at (\xfinal,0) [anchor=west] {};
        \node at (pi,-1) [anchor=north] {\fs $\beta n + \gamma$};
        \node (EndY) at (0,8) [anchor=south] {\fs $[\mean{\pars}]_n$};
        \draw[->] (-0.1,0) -- (EndX);
        \draw[->] (0,-0.2) -- (EndY);
        \foreach \x/\xl in {.5*pi/\tfrac{1}{4}, pi/\tfrac{1}{2},   0.5*3*pi/\tfrac{3}{4}, 2*pi/1 } 
        {
	        \draw (\x,0) node[anchor=north] {\fs $\xl$};
	        \draw (\x,-0.1) -- (\x,0);
	    }
        \foreach \y/\yl in {0/\alpha, .5*pi/\alpha+\tfrac{1}{4}\psi, pi/\alpha+\tfrac{1}{2}\psi,   0.5*3*pi/\alpha+\tfrac{3}{4}\psi, 2*pi/\alpha+\psi}
        {
            \draw (0,\y) node[anchor=east] {\fs $\yl$};
            \draw (-.1*.5*\byxsc,\y) -- (0,\y);
        }
        \draw[color=blue, thick, opacity=0.3] (0,0) -- (2*pi,2*pi) -- (2*pi,0) -- (2*pi+0.1, 0.1);
        \draw[color=blue, dashed, thick, opacity=0.3] (2*pi+0.1, 0.1) -- (\xfinal,\xfinal-2*pi);
         \foreach \p/\n in { \pini/0, 2*pi*\bythree+\pini/1, 4*pi*\bythree+\pini/2}
        {
			\node (point) at (\p,\p) {$\bullet$};
            \node at (point.south) [anchor=north west, yshift=10pt] {\footnotesize $n=\n$};
		}
         
        \foreach \p in { \pini, 2*pi*\bythree+\pini, 4*pi*\bythree+\pini}
        {
            \draw[-stealth,color=olive, very thick, dashed] (\p, \p) -- (\p+\pshift,\p+\pshift);
            \draw[dashed,color=olive] (0,\p+\pshift) -- (\p+\pshift,\p+\pshift);
            \node at (0,\p+\pshift) [color=olive] {\tiny $\blacksquare$};
		}
		 
		\foreach \p in { \pini, 2*pi*\bythree+\pini, 4*pi*\bythree+\pini}
        {
            \draw[-latex,color=red, very thick, dotted] (\p,\p) -- (\p,\p+\pshift);
            \draw[dotted,color=red,thick] (0,\p+\pshift) -- (\p,\p+\pshift);
            \node at (0,\p+\pshift) [color=red] {\tiny $\bullet$};
        }
         
        \draw[dashdotted] (4*pi*\bythree+\pini,0) -- (4*pi*\bythree+\pini,4*pi*\bythree+\pini);
        \draw[dashed,<->, very thick] (4*pi*\bythree+\pini,pi) -- (2*pi,pi);
        \node at (.5*4*pi*\bythree+.5*\pini+pi,pi) [anchor=north] {\small  $\epsilon_+$};
        \draw[dashed,<->,very thick] (pi,4*pi*\bythree+\pini) -- (pi,2*pi);
        \draw[dashdotted] (0,4*pi*\bythree+\pini) -- (4*pi*\bythree+\pini,4*pi*\bythree+\pini);
        \node at (pi,.5*4*pi*\bythree+.5*\pini+pi) [anchor=east] {\small $\psi \epsilon_+$};
\end{tikzpicture}
      
      \vspace{-10pt}
      
    \caption{Examples of non-negative changes in the model parameters $\Delta\alpha\geq0$ and 
			$\Delta\gamma\geq0$ (assuming $\psi>0$) that lead to different (a) or the same (b) 
			means $\mean{\Delta\alpha}$ and $\mean{\Delta\gamma}$ of the data according to model
			\eqref{eq:model} with $\sigma_v=0$ after the respective changes in the model parameters.
			When $\sigma_v=0$, the mean values fully determine identifiability. For any change 
			$\Delta\alpha >0$ (if $\psi<0$, $\Delta\alpha<0$) such that  
			$|\Delta \alpha|\leq \psi \epsilon_+$, there is a positive change in the phase $\Delta\gamma>0$
			that fulfills $\Delta \gamma \leq \epsilon_+$ and yields the same mean. 
			The same is true vice versa. \label{fig:unidentifiable}
    }
  \end{figure}
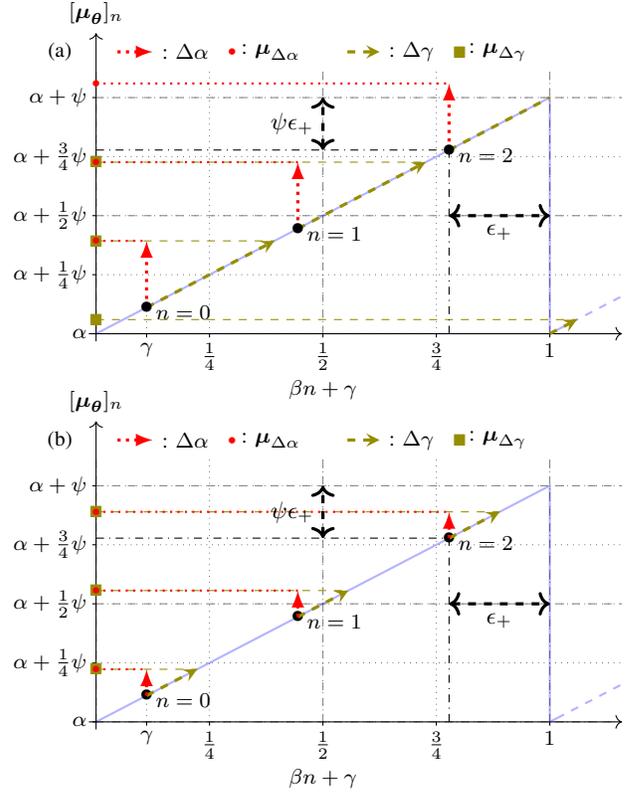
  
  Consider first \eqref{eq:model} with $\sigma_v=0$ and observe that then, $\model =
    \normal{\mean{\pars}}{\sigma_w^2 \Id{N}}$, where $\Id{N}$ is the $N\times N$ 
  identity matrix and
  \begin{IEEEeqnarray}{c}\label{eq:mean}
    \mean{\pars} = \alpha \ones{N}+ \psi \modu\left(
      \beta \mathbf{n} + \gamma\ones{N}
      \right)\,,
  \end{IEEEeqnarray}
  with the modulus operation $\modu(\cdot)$ applied component-wise, 
  $\ones{N}=\trans{[1,1,\dots,1]}\in\reals^N$, and 
  $\mathbf{n}=\trans{[0,1,\dots,N-1]}$. Because the normal distribution is fully characterized by
  its location and scale parameters, we know that changes in $\pars$ will only
  affect the distribution in terms of its location, controlled by its mean
  $\mean{\pars}$. Consequently, the condition for identifiability in Def.~\ref{def:id}, i.e., \eqref{eq:def:id}, 
  can be restated as $\mean{\parsone} = \mean{\parstwo} \Leftrightarrow \parsone = \parstwo$. 
  Lem.~\ref{lem:unid-nojit} establishes that, when $\sigma_v=0$, there are 
  changes in $\alpha$ and $\gamma$ that violate this condition.	
  
  \begin{lemma}[Unidentifiability of \eqref{eq:model} when $\sigma_v=0$]
  \label{lem:unid-nojit}
      Let $\model$ express the model of the data $\datarv=\trans{\left[ Y[0], Y[1], \dots, Y[N-1] \right]}$
      given by \eqref{eq:model} with $\pars = \trans{[\alpha,\psi,\beta,\gamma]}$, $\alpha\in\reals$, $|\psi|>0$, 
      $\beta \in [-1/2,1/2)$ and $\gamma \in [0,1)$, when the inner noise is disregarded, i.e.,  $\sigma_v=0$. 
      Then, $\model$ is unidentifiable. In particular, there are different combinations of $\alpha$ and $\gamma$ 
      that yield the same distribution of $\datarv$ under $\model$.
  \end{lemma}
  \begin{IEEEproof}
    We will show that $\mean{\parsone} = \mean{\parstwo} \not\Rightarrow \parsone = \parstwo$, i.e., the forward implication of
    \eqref{eq:def:id} in Def.~\ref{def:id} is not fulfilled. In particular, given a parameter vector $\pars=\trans{[\alpha,\psi,\beta,\gamma]}$,
    we will find $\parsone,\parstwo$ such that $\parsone \neq \parstwo$ and $\mean{\parsone} = \mean{\parstwo}$.
    
    Observe that, because $\modu:\reals\rightarrow[0,1)$, 
    \begin{IEEEeqnarray*}{c} 
      \epsilon_+ = 1-\max{n\in\listintzero{N-1}}{\modu\left(
      \beta n + \gamma
      \right)} > 0\,.
    \end{IEEEeqnarray*}
    Then, $\forall n \in \listintzero{N-1}$ and $\forall \epsilon\in[0,\epsilon_+)$,
    \begin{IEEEeqnarray*}{c}
      \alpha + 
      \psi \modu\left( \beta n + \left[ \gamma + \epsilon \right] \right) 
      = \left[ \alpha + \psi \epsilon \right] +
      \psi \modu\left( \beta n + \gamma \right)\,.
    \end{IEEEeqnarray*}
    Therefore, for any $\epsilon\in[0,\epsilon_+)$,
    \begin{IEEEeqnarray*}{c}
      \parsone = \trans{\left[ \alpha + \psi\epsilon, \psi, \beta, \gamma \right]} \mbox{ and }
      \parstwo = \trans{\left[ \alpha, \psi, \beta, \gamma + \epsilon  \right]}\,,
    \end{IEEEeqnarray*}
    yield $\mean{\parsone} = \mean{\parstwo}$, i.e., 
    $\mean{\parsone} = \mean{\parstwo} \not\Rightarrow \parsone = \parstwo$ and
    $\model$ is unidentifiable.
    
  \end{IEEEproof}
  Fig.~\ref{fig:unidentifiable} illustrates the idea of our proof of Lem.~\ref{lem:unid-nojit} when 
  $\psi>0$, and shows, in Fig.~\ref{fig:unidentifiable}(b), changes in $\alpha$ and $\gamma$ that cannot
  be distinguished in the mean of $\datarv$ for a simple example with $N=3$.
  Consequently, Fig.~\ref{fig:unidentifiable}(b) serves as a straightforward counter-example to the 
  identifiability of $\model$ when $\sigma_v=0$. Note that both the result in Lem~\ref{lem:unid-nojit} and the counter-example of Fig.~\ref{fig:unidentifiable}(b) are valid when $\sigma_w=0$, i.e., when the 
  model is deterministic as in \eqref{eq:ideal_model}. In contrast with our result in Theorem~\ref{th:id}, this implies arbitrarily accurate synchronization could be impossible in an idealized scenario without noise.
  
  In our proof of Lem.~\ref{lem:unid-nojit}, we exploit the formal definition of $\modu(\cdot)$ to 
  claim that its value will always be strictly less than one, and therefore, we obtain the margin 
  $\epsilon_+$ under which changes in $\alpha$ of the same sign as $\psi$ and positive changes in  
  $\gamma$ are not distinguishable. However, the real limitation on identifiability is given by the 
  points closest to the discontinuity from both sides, and, in most cases (i.e., $\gamma\neq 0$ and
  for most $\beta$s), a similar margin $\epsilon_{-}$ can be obtained under which changes in $\alpha$
  of the sign opposite to $\psi$ and negative changes in $\gamma$ are not distinguishable.
  
    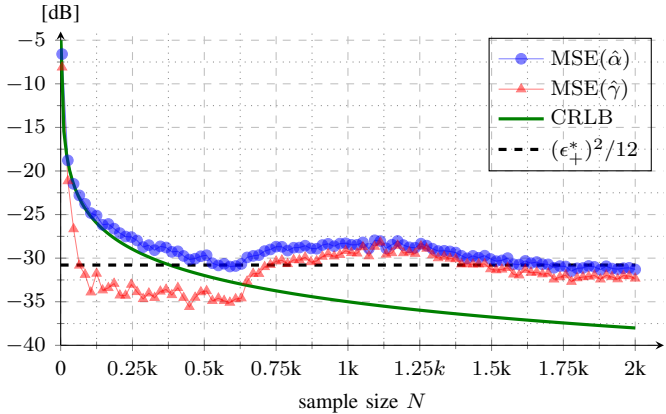
\begin{figure}
    \centering
    \begin{tikzpicture}
\begin{axis}[
    width = 1.084\linewidth,
    height = .65\linewidth,
    legend entries = { {$\mathrm{MSE}(\hat{\alpha})$}, {$\mathrm{MSE}(\hat{\gamma})$},
                       {$\mathrm{CRLB}$}, {$(\epsilon^*_+)^2 / 12$} },
    legend style = {black,fill=none},
    legend cell align={left},
    grid = both,
    major grid style={dashed,gray!50},
    minor grid style={dotted,gray!90},
    minor tick num = {1},
    axis x line = bottom,
    axis y line = left,
    y label style = {at={(0,1)},rotate = -90,anchor=south},
    ylabel={[dB]},
    xlabel = {sample size $N$},
    xtick = {0,250,...,2000},
    xticklabels={0,$0.25$k, $0.5$k,$0.75$k, $1$k, $1.25k$, $1.5$k, $1.75$k, $2$k},
    ytick = {-40,-35,...,-5},
    xmin = 0, xmax=2099.75,
    ymin=-40, ymax=-4,
]
\addplot [blue, mark=*, mark size=2, mark options={opacity=0.5}, opacity = 0.5]
table [row sep=\\]{%
5	-6.58874694903914 \\
25	-18.7842755760075 \\
45	-21.4810423200899 \\
65	-22.8004574413137 \\
85	-23.769693956325 \\
105	-24.8274557750159 \\
125	-25.1206909834352 \\
146	-26.1742680347443 \\
166	-26.0775192213745 \\
186	-26.616791312787 \\
206	-27.1287509137639 \\
226	-27.5186095138873 \\
246	-27.5836667985421 \\
266	-27.9671937495973 \\
287	-28.7334770830376 \\
307	-28.5064608998965 \\
327	-29.0846621245033 \\
347	-28.6447046779866 \\
367	-28.7773869942602 \\
387	-29.2780940057941 \\
408	-29.2149143323929 \\
428	-29.7519750356968 \\
448	-30.1625478668514 \\
468	-30.0615360399943 \\
488	-29.7793957558841 \\
508	-29.6641903187513 \\
528	-30.5809688041723 \\
549	-30.8341118795238 \\
569	-30.6445403887255 \\
589	-30.9624552733865 \\
609	-30.9004937565002 \\
629	-30.6673639594936 \\
649	-29.9144113345202 \\
670	-29.7413976836875 \\
690	-29.6517161391916 \\
710	-29.1885416278333 \\
730	-29.0029899842336 \\
750	-28.6813462945649 \\
770	-28.695460051145 \\
790	-29.1314817564997 \\
811	-28.8023621438441 \\
831	-28.7854004735718 \\
851	-28.5691214889351 \\
871	-28.8847753187958 \\
891	-28.9212830935613 \\
911	-28.4958760083891 \\
931	-28.3548334317251 \\
952	-28.6324348994611 \\
972	-28.3253068823061 \\
992	-28.2892472990929 \\
1012	-28.4673443203836 \\
1032	-28.4658628627206 \\
1052	-28.1889910039293 \\
1073	-28.9029266249425 \\
1093	-27.9484271280857 \\
1113	-28.0422111306098 \\
1133	-28.8722086791465 \\
1153	-28.4559024825095 \\
1173	-28.0281482528865 \\
1193	-28.5004492422429 \\
1214	-28.7545178084564 \\
1234	-28.8012880441435 \\
1254	-28.336069023996 \\
1274	-28.5624996834858 \\
1294	-28.9149572455299 \\
1314	-29.1812201777255 \\
1335	-29.1537298387996 \\
1355	-29.2269342694603 \\
1375	-29.3545463724399 \\
1395	-29.7251973852499 \\
1415	-30.0190159402497 \\
1435	-29.7370818949362 \\
1455	-29.9524658076731 \\
1476	-29.7138657194819 \\
1496	-30.0894219656806 \\
1516	-30.3463306165171 \\
1536	-30.2570141535263 \\
1556	-29.9957995756059 \\
1576	-30.6955040163216 \\
1596	-30.7366464906587 \\
1617	-30.3198828437723 \\
1637	-30.5768186354722 \\
1657	-30.4456913028505 \\
1677	-30.7032780418636 \\
1697	-30.7115442597675 \\
1717	-31.1076146292636 \\
1738	-30.9126378929251 \\
1758	-30.9579796877311 \\
1778	-31.4877428104996 \\
1798	-31.1652865208219 \\
1818	-31.129252988104 \\
1838	-30.8955187823577 \\
1858	-30.9997520784785 \\
1879	-31.3163187105829 \\
1899	-30.9257924441866 \\
1919	-31.1953466757227 \\
1939	-31.2601539277768 \\
1959	-31.1514615671037 \\
1979	-31.1070632688789 \\
2000	-31.2975293534516 \\
};

\addplot [red, mark=triangle*, mark size=2, mark options={opacity=0.5}, opacity=0.5]
table [row sep=\\]{%
5	-8.11133110385787 \\
25	-21.1511397347707 \\
45	-26.6587683329657 \\
65	-30.8324796584923 \\
85	-31.8927256894208 \\
105	-33.9105357335187 \\
125	-31.7596858845528 \\
146	-33.6753735486998 \\
166	-33.412272219488 \\
186	-32.9207716260124 \\
206	-34.2620341785126 \\
226	-34.284692559374 \\
246	-33.0258169873189 \\
266	-33.8726007263914 \\
287	-34.6736022415921 \\
307	-34.0309340508873 \\
327	-34.5443622412385 \\
347	-33.813468141763 \\
367	-33.5987565745984 \\
387	-34.2197584807559 \\
408	-33.4050480092133 \\
428	-34.5355634850656 \\
448	-35.5836918295408 \\
468	-34.5232791627649 \\
488	-33.890998216194 \\
508	-33.7633987087616 \\
528	-34.8922041796174 \\
549	-34.6064247451432 \\
569	-34.8880462883939 \\
589	-35.0913189239131 \\
609	-34.6890840133268 \\
629	-34.5923906456837 \\
649	-32.669427343114 \\
670	-31.773449548101 \\
690	-31.7830626941519 \\
710	-31.0038960659061 \\
730	-30.7578062761649 \\
750	-30.3591052695987 \\
770	-30.2483545406815 \\
790	-30.6583357336164 \\
811	-30.2006877903516 \\
831	-30.2196698950351 \\
851	-29.9137023794064 \\
871	-30.4436803772813 \\
891	-30.3267343430782 \\
911	-29.8159273359386 \\
931	-29.5268934835744 \\
952	-29.7850076737457 \\
972	-29.6143453926825 \\
992	-29.1575184823735 \\
1012	-29.3929412617007 \\
1032	-29.4465409456134 \\
1052	-28.7227252773959 \\
1073	-29.7674035043788 \\
1093	-28.6687666586434 \\
1113	-28.1926758843532 \\
1133	-29.5733116784752 \\
1153	-29.405132900653 \\
1173	-28.5542687510878 \\
1193	-29.4106702703231 \\
1214	-29.3059611461363 \\
1234	-29.4828585425844 \\
1254	-28.6326677450932 \\
1274	-28.8970342642631 \\
1294	-29.6448624250984 \\
1314	-29.9901464699533 \\
1335	-29.8205578506226 \\
1355	-30.034006121067 \\
1375	-30.1021001947338 \\
1395	-30.6927706317285 \\
1415	-30.6625645898409 \\
1435	-30.7124101542549 \\
1455	-30.6944880198046 \\
1476	-30.6710027733659 \\
1496	-31.2505794631791 \\
1516	-31.3719479936957 \\
1536	-31.2891936293968 \\
1556	-31.0967700480333 \\
1576	-31.8566713550573 \\
1596	-31.9585115530467 \\
1617	-31.3208699542782 \\
1637	-31.6525210055608 \\
1657	-31.6218130735556 \\
1677	-31.8396621416022 \\
1697	-31.7514478209606 \\
1717	-32.4282285553367 \\
1738	-32.1683705779778 \\
1758	-31.9479062570264 \\
1778	-32.6941519253933 \\
1798	-32.2849361868828 \\
1818	-32.1686267462138 \\
1838	-32.1087906672293 \\
1858	-31.9877400457159 \\
1879	-32.4263256210519 \\
1899	-31.9794163552919 \\
1919	-32.1553183507422 \\
1939	-32.405727165747 \\
1959	-32.0062646385288 \\
1979	-32.0566860744755 \\
2000	-32.3225915900332 \\
};

\addplot [very thick, green!50!black, domain = 1:2000, samples = 200] {-5 - 10*log10(x)};

\addplot [very thick, black, dashed, mark = none, samples = 2, domain = 0:2000] {-30.7918124604763};

\end{axis}

\end{tikzpicture}

    \vspace{-20pt}

    \caption{$\mathrm{MSE}(\hat{\alpha}_\mathrm{GGS})$ and $\mathrm{MSE}(\hat{\gamma}_\mathrm{GGS})$
        against the sample 
        size $N$ when $\sigma_v=0$, $\psi = 1$ and $\beta=M/Q$ with $M=1$ and $Q=10$. $\hat{\alpha}_\mathrm{GGS}$ and 
        $\hat{\gamma}_\mathrm{GGS}$ are the result of a global grid search (GGS) on the prediction mean squared error 
        with model \eqref{eq:ideal_model} (see~\cite{AguilaPla2020}) with $1000$ discretization points for 
        $\gamma \in[0,1)$ when $\beta$ and $\psi$ are known. 
        Results obtained from $2000$ Monte Carlo repetitions and $\SNRout = 5~\mathrm{dB}$ (see Table~\ref{tab:physical_and_sim_parameters}).
        To access the fully reproducible code to generate this figure, see this project's 
        repository~\cite{GITHUB}. For comparison, the figure shows the Cram\'er-Rao lower bounds (CRLB) for estimating 
        an offset in white noise, which
        here serves as a reference for the estimation of $\alpha$ and $\gamma$ (see the supplementary material of this paper
        for details). Here, we observe that while the MSE in estimating $\alpha$ initially decays as predicted by the 
        CRLB, it stabilizes around the variance of a uniform noise of width $\epsilon^*_+$. Furthermore,
        the MSE in estimating $\gamma$ becomes worse and stabilizes around the same value when the estimation of $\alpha$
        reaches that level. \label{fig:epsilon_phenomenon}}
    \end{figure}

  While our analysis is concerned with a fixed value of $N$, the lack of identifiability stated in 
  Lem.~\ref{lem:unid-nojit} may be less problematic in an asymptotic regime. 
  In particular, if increasing the sample size tends to reduce the segment at the left of the 
  non-linearity without any sample, i.e., $\epsilon_+\rightarrow 0$ when $N\rightarrow+\infty$,
  the size of the changes in $\alpha$ and $\gamma$ that cannot be distinguished in the data would 
  decrease with $N$, making the model identifiable in an asymptotic regime, or at least taking away importance from our proof of non-identifiability for large $N$s. 
  In particular, if we consider the sequence of elements of the vector \eqref{eq:mean}, ${\mean{\pars}}_{n}$, as a 
  sequence, we obtain what is known in dynamical systems as the orbit of a rotation of the circle.
  Then, if $\beta\in\reals\setminus\rationals$ we have that the orbit is 
  minimal~\cite[ch. 1.3., proposition 1.3.3.]{Katok1995}, i.e., that the set
  $\lbrace{\mean{\pars}}_{n}\rbrace_1^N$ when 
  $N\rightarrow+\infty$ is dense in $[0,1)$, and thus, $\epsilon_+\rightarrow 0$.  
  This contradicts the intuitive notion of finite-sample identifiability as a necessary condition
  for consistent estimators to exist, seen, for example, in \cite[p.~62]{Vaart1998}.
  In contrast, when $\beta \in \rationals$, \eqref{eq:mean} is periodic and hence 
  $\epsilon_+ \rightarrow \epsilon_+^* > 0$. In particular, if $\beta = \pm M/Q$ with $M$ and $Q$ 
  two co-prime naturals, then \eqref{eq:mean} is periodic with minimal period $Q$, and increasing the 
  sample size beyond $Q$ will not result in any further reduction of $\epsilon_+$, i.e., any further
  improvement from an identifiability perspective. In the case of 
  clock synchronization, this specific case corresponds to coherent sampling, in which 
  $\samplingperiod\freqd = \pm M/Q$. The effect of this specific case in the estimation error of a 
  global grid search (GGS) strategy on the prediction mean square error of the model \eqref{eq:ideal_model}
  (see~\cite{AguilaPla2020}) when $\psi$ and $\beta$ are known is illustrated in Fig.~\ref{fig:epsilon_phenomenon}.

  Our analysis has assumed that $\alpha$ was part of the parameter vector $\pars$, and that 
  one wanted to recover it. Although this can be the prominent case in many applications of 
  the sawtooth model, e.g., synchronization in wireless sensor networks or networks of 
  autonomous vehicles, other applications may consider $\alpha$ to be known.
  Within synchronization, this would be the case of base-station synchronization in cellular 
  networks, in which the backhaul links will most likely have a known and stable transmission
  delay. This would invalidate the identifiability analysis in Lem.~\ref{lem:unid-nojit}, and under
  some additional conditions, $\model$ could be shown to be identifiable. Regardless, in the 
  next section we analyze the full model in the presence of noise inside the $\mod{1}\left(\cdot\right)$ 
  non-linearity, i.e., with $\pars = \trans{[\alpha,\psi,\beta,\gamma]}$ when $\sigma_v>0$, 
  and show its identifiability.

  \subsection{Identifiability with inner noise} \label{sec:id_all}
      \begin{theorem}[Identifiability of \eqref{eq:model} when $\sigma_v>0$] \label{th:id}
    Let $\model$ express the model of the data $\datarv = \trans{\left[ Y[0], Y[1], \dots, Y[N-1] \right]}$
    given by \eqref{eq:model} when $\pars = \trans{[\alpha,\psi,\beta,\gamma]}$, the parameters fulfill 
    $\alpha\in\reals$, $|\psi|>0$ with $\mathrm{sign}(\psi)$ known, $\beta \in [-1/2,1/2)$ and $\gamma \in [0,1)$,
    there is noise inside the $\mod{1}\left(\cdot\right)$ non-linearity, i.e., $\sigma_v>0$ and $\sigma_w\geq0$,
    and at least two RTT measurements have been taken, i.e., $N\geq 2$. 
    Then, $\model$ is an identifiable model and $\pars$ is an identifiable parameter.
  \end{theorem}
  
  Because the proof of Theorem~\ref{th:id} is rather technical, we place it in the Appendix~\ref{app:id_proof}. However, it is worthwhile to note here
  that it is not limited to the case in which $W[n]$ and $V[n]$ are Gaussian processes. Indeed, the statements in there apply \emph{mutatis mutandis} 
  under a wide variety of distributions for $W[n]$ and $V[n]$. In particular, any $W[n]$ consisting of independent and identically distributed (IID)
  samples from any location-scale family with some reference PDF $\varphi\!\left(w\right)$ with unbounded support will allow for the conclusion in 
  \eqref{eq:rhos}. Furthermore, such a $W[n]$ together with any $V[n]$ consisting of IID samples from a location family that accepts a PDF and leads 
  to a monomodal distribution after wrapping with mode equal to the location parameter, e.g., IID Cauchy distributed samples \cite[p.~51]{Mardia2009}, 
  will also preserve all the statements therein. Nonetheless, to our knowledge, the literature mostly considers timing errors to be Gaussian (see, among others, 
  \cite{Noh2007,Skog2010,Zheng2010,Chepuri2013,Koivisto2017,Zachariah2017,Geng2018}), with little empirical incentive to consider other models.

  The contrast between Lem.~\ref{lem:unid-nojit} and Theorem~\ref{th:id} is initially non-intuitive. Indeed, it implies that the presence of noise inside the 
  non-linearity improves the theoretical condition of the estimation problem. This result recalls the popular theories of stochastic resonance for testing 
  and estimation \cite{Kay2000,Chen2007,Chen2008,Kay2008,Chen2014} and of dithering for improving the signal-independence of quantization noise \cite{Schuchman1964}, 
  but is, in fact, less expected. In summary, both these theories delve into using noise to improve the performance of knowingly suboptimal strategies. In contrast, our 
  identifiability result reveals how the inclusion of noise makes the data more informative with respect to the underlying parameters. To understand this result, one must first consider that it is very specific to sawtooth models, as it relies on that for any $x\in\reals\setminus\integ$, there is an $\epsilon>0$ such that $\mod{1}(x+\epsilon) =  \epsilon + \mod{1}(x)$. In other words, small changes in phase and offset are equivalent around almost every $x$. In order to have any hope to distinguish them, one needs to guarantee that the points in which their effect differs play a role in shaping the distribution of the data. A phase noise $V[n]$ with long enough tails provides this guarantee, because changing the phase will alter the wrapping of the tails of the distribution through these non-linear points, while changing the offset will not.



\section{Numerical results} \label{sec:numres}

  	\begin{table}
		\caption{Values for the physical and simulation parameters in our numerical results, unless otherwise stated.
			\label{tab:physical_and_sim_parameters}}
		\centering
		\begin{tabular}{c|c|c}
				Parameter & Interpretation & Value \\ \hline
				$N$ & 
				sample size &
				$2000$ \\
				$\delay$ & 
				delay introduced by $\slave~[\mathrm{s}]$ &
				$5\cdot 10^{-6}$ \\
				$\masterperiod$ &
				$\master$'s period $[\mathrm{s}]$ & 
				$10^{-8}$ \\
				$\samplingperiod$ & 
				sampling period $[\mathrm{s}]$ &
				$10^{-4}$ \\
				$\rho$ &
				range $[\mathrm{m}]$ &
				$\uniform{\left[1}{3\right)}$ \\
				$\freqd$ &
				frequency difference $[\mathrm{Hz}]$ & 
				$\uniform{\left[-200}{200\right)}$ \\
				$\slavephase$ & 
				$\slave$'s phase $[\mathrm{rad}]$ &
				$\uniform{\left[0}{2\pi\right)}$ \\
				$\SNRin$ &
				SNR for $V[n]$, $ 1 / \sigma_v^2~[\mathrm{dB}]$ &
				$40$ \\
				$\SNRout$ & 
				SNR for $W[n]$, $\psi^2 / \sigma_w^2~[\mathrm{dB}]$ &
				$20$ \\\hline
		\end{tabular}	
	\end{table}
        
    In~\cite{AguilaPla2020}, we propose two estimators for the parameters of the 
    sawtooth model \eqref{eq:model}. In particular, we first introduce an heuristic technique
    that is computationally light, intuitive, and surprisingly robust, which we call 
    periodogram and correlation peaks (PCP). In this approach, one uses peaks in the
    power of the discrete Fourier transform to estimate the normalized frequency, and peaks in a circular 
    correlation of the first estimated period to estimate the phase parameter.
    Using PCP as a starting point, we also introduce the local grid search (LGS), a computationally heavy
    method that improves the final performance by exploring a grid of normalized frequencies and 
    phase parameters around the PCP estimate, and picks the combination that minimizes the 
    model's prediction mean square error, i.e., how well the model fits the observed signal.
    In both techniques, the offset parameter is chosen as the least squares estimate for each 
    pair of frequency and phase estimates. For more detail on these techniques, and for the
    values of their parameters in the experimental results presented here,
    see~\cite{AguilaPla2020,GITHUB}.
    
	Here, we evaluate these estimators in a series of realistic simulation studies
	of clock synchronization and ranging. In the exposition of these 
	results we intend to aid 1) engineers that aim to
	apply this technology, by providing reasonable expectations on its current possibilities, and 2) theoreticians that aim to develop estimators
	for the parameters of the sawtooth signal model \eqref{eq:model}, by revealing the strengths and pitfalls of the techniques that are currently 
	available. With respect to 1), we include in all our results references that make it easier to identify different standard performance 
	measures, regardless of the scale and aspect of each figure. In particular, a) in figures reporting ranging performance, i.e., 
	$\mse{\hat{\rho}}$, horizontal lines corresponding to standard errors of 
	$1~\mathrm{cm}$ or 
	$0.1~\mathrm{cm}$  
	are shown, 
	b) in figures reporting frequency-difference estimation performance, i.e., $\mse{\hat{\freqd}}$, horizontal lines corresponding to 
	standard errors of 
	$10~\mathrm{ppb}$ ($1~\mathrm{Hz}$) and 
	$1~\mathrm{ppb}$ ($0.1~\mathrm{Hz}$) 
	of $\master$'s frequency, 
	$1/\masterperiod = 100~\mathrm{MHz}$ (see Table~\ref{tab:physical_and_sim_parameters}), are shown, and 
	c) in figures reporting $\slave$'s phase estimation performance, i.e., $\mse{\hat{\slavephase}}$, horizontal lines corresponding to
	standard errors in $\varphi_\slave$ that are one 
	or two orders of magnitude 
	below $\slaveperiod$ are shown. 
	In reference to c), we intentionally report performance on the estimation of the phase parameter $\slavephase$ instead of the absolute 
	time delay $\varphi_\slave = \slaveperiod \slavephase / (2\pi)$. In our opinion, this is a better and fairer measure of how useful a 
	specific clock synchronization technique is, because the scale of the errors in $\varphi_\slave$ will always be dominated by the order
	of magnitude of $\slaveperiod$. In other words, if $\slaveperiod \approx 10~\mathrm{ns}$, even guessing $\slavephase$ at random between 
	$0$ and $2\pi$ achieves errors in $\varphi_\slave$ that are on the order of $\mathrm{ns}$. In order to streamline the exposition of 
	this section and avoid unnecessary repetitions of the experimental conditions, we detail the default values for the physical and simulation
	parameters in Table~\ref{tab:physical_and_sim_parameters}. Note here that we chose to randomize 
	the most critical parameters, so that the dependences shown in 
	Figs.~\ref{fig:vs_freq_d_randomized}--\ref{fig:vs_SNRs_randomized} can be considered the average
	behavior within the selected ranges.
    
	\subsection{Sensitivity to the parameter values}

		One of the weaknesses of the estimation approaches we present in~\cite{AguilaPla2020} is
		low performance when $\freqd$ is small. In particular, because the PCP includes a mean-removal step
		before the DFT, the low frequencies are supressed. If $\freqd$ is small, then,
		the peak we aim to detect in the DFT is most likely dampened and we detect noise instead.
		This is visualized in Fig.~\ref{fig:vs_freq_d_randomized}, which shows the average performance
		in the estimation of $\freqd$ by both PCP and LGS as a function of $\freqd$, when all other
		parameters are randomized according to Table~\ref{tab:physical_and_sim_parameters}.
		Here, we see that the estimators fail when $\freqd\in[-10,10)$. 
		Furthermore, we also observe the effect of the grid underlying the DFT on the PCP frequency 
		estimate. In particular, for a given $N$, the PCP will only propose as estimates frequencies 
		that are at one of the points in the DFT (e.g., $-200~\mathrm{Hz}$ and $200~\mathrm{Hz}$
		in Fig.~\ref{fig:vs_freq_d_randomized}), and so frequencies that are close to those will be 
		better estimated than those that are far. This same phenomenon is observed in the experimental
		results we present in~\cite{AguilaPla2020} for varying $N$ and a fixed combination of physical
		parameters.
		Remarkably, the performance obtained with 
		LGS remains below $1~\mathrm{ppb}$ of $1/\masterperiod$ for most frequencies $\freqd$ outside the 
		$[-10,10)$ area.
    
    \begin{figure}
			\centering
			\input{figs/vs_f_d_randomized}
			
			\vspace{-20pt}
			
			\caption{ Result of $300$ Monte Carlo repetitions in the conditions specified 
					in Table~\ref{tab:physical_and_sim_parameters} evaluating the MSE in the
					estimation of $\freqd$ by both PCP and LGS with respect to its actual
					value. \label{fig:vs_freq_d_randomized} }
		\end{figure}
		
		A weakness of the measurement protocol described in Section~\ref{sec:ideal_model} is that, due to
		the lack of time-stamps in the exchanged packets, the time origin shift between $\master$ and 
		$\slave$'s clock appears only in terms of a phase term inside the $\modu(\cdot)$ function, 
		e.g., $\slavephase$ if we assume that $\varphi_\master=0$. If one desires absolute time
		synchronization, this can have dire consequences. While $\slavephase=2\pi(1-\xi/2)$ and 
		$\slavephase=\pi\xi$ for some small $\xi>0$ are only $2\pi\xi$ away under the non-linear 
		wrapping, their difference implies errors in $\varphi_\slave$ of $\slaveperiod(1-\xi)$. 
		In order to take this effect into account, we use the conventional Euclidean distance to 
		quantify the error for the phase term $\slavephase$, without taking into account the wrapping 
		effect. The ensuing increase of error around the extremes can be seen plainly 
		in Fig.~\ref{fig:vs_phase_s_randomized}. This weakness is characteristic of protocols for
		clock synchronization without time-stamping. In turn, however, these protocols
		minimize the communication overhead and are more robust to malicious nodes \cite[p.~29]{Etzlinger2018}.
		
		\begin{figure}
				\centering
				\begin{tikzpicture}

\begin{axis}[
    width = 1.07\columnwidth,
    height = .6\columnwidth,
    legend entries ={{$\mse{ \hat{\slavephase}_{\mathrm{PCP}} }$},{$\mse{ \hat{\slavephase}_{\mathrm{LGS}} }$},{$(2\pi/10)^2$},{$(2\pi/100)^2$}},
    legend style={at={(0.5,0.95)}, anchor=north, draw=black, fill=none},
    legend columns = 2,
    legend cell align={left},
    grid = both,
    major grid style={dashed,gray!50},
    minor grid style={dotted,gray!90},
    minor tick num = {1},
    axis x line = bottom,
    axis y line = left,
    y label style = {at={(0.05,1)},rotate = -90,anchor=south},
    ylabel={$\mathrm{rad}^2$ [dB]},
    x label style = {at={(1,0.025)},anchor = south east},
    xlabel = {$\slavephase~[\mathrm{rad}]$},
    xtick={0,0.785398163397448,1.5707963267949,2.35619449019234,3.14159265358979,3.92699081698724,4.71238898038469,5.49778714378214,6.28318530717959},
    xticklabels={$0$,$\frac{\pi}{4}$,$\frac{\pi}{2}$,$\frac{3}{4}\pi$,$\pi$,$\frac{5}{4}\pi$,$\frac{3}{2}\pi$,$\frac{7}{4}\pi$,$2\pi$},
    ytick={-25, -20, ..., 15},
    xmin=-0.05, xmax=6.41,
    ymin=-26, ymax=16.4,
]
\addplot [blue, mark=*, mark size=2, mark options={opacity = 0.5},opacity=0.5]
table [row sep=\\]{%
0	14.1047632792209 \\
0.0634665182543393	10.6953109521939 \\
0.126933036508679	8.19090071863282 \\
0.190399554763018	4.51317714085396 \\
0.253866073017357	1.11234132066183 \\
0.317332591271696	-3.11048222729704 \\
0.380799109526036	-9.21721160842677 \\
0.444265627780375	-6.37855892948981 \\
0.507732146034714	-7.44555147973957 \\
0.571198664289053	-9.88703037688149 \\
0.634665182543393	-3.68266414955598 \\
0.698131700797732	-2.80966928825215 \\
0.761598219052071	-3.5939448012472 \\
0.82506473730641	-3.66238002658087 \\
0.88853125556075	-4.12196498609228 \\
0.951997773815089	-3.69597611951101 \\
1.01546429206943	-9.27742356881253 \\
1.07893081032377	-11.2729026142438 \\
1.14239732857811	-4.23980650141265 \\
1.20586384683245	-6.49459704269222 \\
1.26933036508679	-9.94480593117013 \\
1.33279688334112	-8.26047421120811 \\
1.39626340159546	-5.79215419213467 \\
1.4597299198498	-7.85398101316329 \\
1.52319643810414	-7.57989704655398 \\
1.58666295635848	-8.97161022394783 \\
1.65012947461282	-9.19379320866571 \\
1.71359599286716	-9.74454636626377 \\
1.7770625111215	-11.4708166954064 \\
1.84052902937584	-10.587653417902 \\
1.90399554763018	-5.91712564014747 \\
1.96746206588452	-13.7871300965687 \\
2.03092858413886	-6.51049880877491 \\
2.0943951023932	-10.4682672310853 \\
2.15786162064753	-8.93608403817831 \\
2.22132813890187	-8.10141881569478 \\
2.28479465715621	-10.8981876382205 \\
2.34826117541055	-11.7144721124667 \\
2.41172769366489	-8.16183692560603 \\
2.47519421191923	-10.2663069596562 \\
2.53866073017357	-11.5634980983786 \\
2.60212724842791	-10.2740389180387 \\
2.66559376668225	-11.0920122774378 \\
2.72906028493659	-11.6828422431217 \\
2.79252680319093	-10.5835407709453 \\
2.85599332144527	-18.8600834801846 \\
2.91945983969961	-9.75835770789529 \\
2.98292635795395	-11.1809350588311 \\
3.04639287620828	-9.98634522260581 \\
3.10985939446262	-10.2283199756683 \\
3.17332591271696	-8.57774626510788 \\
3.2367924309713	-8.60220691003184 \\
3.30025894922564	-12.3956861006704 \\
3.36372546747998	-10.2029770852818 \\
3.42719198573432	-9.36799237876636 \\
3.49065850398866	-10.88459517269 \\
3.554125022243	-7.76188354330099 \\
3.61759154049734	-15.183053483832 \\
3.68105805875168	-13.9348492354835 \\
3.74452457700602	-8.76291686461512 \\
3.80799109526036	-15.9256022523515 \\
3.87145761351469	-14.3927142662385 \\
3.93492413176903	-15.4087566994805 \\
3.99839065002337	-13.3974047200972 \\
4.06185716827771	-10.8386224021566 \\
4.12532368653205	-13.6290549190071 \\
4.18879020478639	-11.2003063152105 \\
4.25225672304073	-7.65437884007639 \\
4.31572324129507	-9.62150160612351 \\
4.37918975954941	-10.1227298307612 \\
4.44265627780375	-13.0667180800712 \\
4.50612279605809	-12.2782479515929 \\
4.56958931431243	-9.82293905100952 \\
4.63305583256677	-10.6661045728554 \\
4.69652235082111	-9.83083437434768 \\
4.75998886907544	-6.7398416194643 \\
4.82345538732978	-10.9234020160228 \\
4.88692190558412	-8.24077039937339 \\
4.95038842383846	-9.56484886655353 \\
5.0138549420928	-11.3872121417365 \\
5.07732146034714	-6.03096163021395 \\
5.14078797860148	-7.26551685535285 \\
5.20425449685582	-9.53458808445582 \\
5.26772101511016	-6.23868963619685 \\
5.3311875333645	-8.62447542908085 \\
5.39465405161884	-5.89124611052289 \\
5.45812056987318	-6.2100528486315 \\
5.52158708812752	-9.3738774573792 \\
5.58505360638185	-10.213894399067 \\
5.64852012463619	-7.07152562950677 \\
5.71198664289053	-5.41866145175414 \\
5.77545316114487	-7.90583710001095 \\
5.83891967939921	-8.25617985181522 \\
5.90238619765355	-3.72372042368174 \\
5.96585271590789	-4.49566213702975 \\
6.02931923416223	-11.73365766402 \\
6.09278575241657	-3.80760941010572 \\
6.15625227067091	-0.223881996750477 \\
6.21971878892525	6.59029104216321 \\
6.28318530717959	11.8009011557203 \\
};
\addplot [orange, mark=triangle*, mark size=2, mark options={opacity = 0.5},opacity = 0.5]
table [row sep=\\]{%
0	13.572933592062 \\
0.0634665182543393	5.19852504363187 \\
0.126933036508679	0.57114712315087 \\
0.190399554763018	-2.37626803587253 \\
0.253866073017357	-0.677553927629792 \\
0.317332591271696	-5.66427283818636 \\
0.380799109526036	-8.3487507885033 \\
0.444265627780375	-6.98842116209376 \\
0.507732146034714	-4.77038423477316 \\
0.571198664289053	-8.70339572812499 \\
0.634665182543393	-4.12425979242466 \\
0.698131700797732	-2.47256390154453 \\
0.761598219052071	-3.1069950851983 \\
0.82506473730641	-3.17516397422445 \\
0.88853125556075	-2.98528846975157 \\
0.951997773815089	-4.96823971688127 \\
1.01546429206943	-12.5052780100583 \\
1.07893081032377	-8.46302700361703 \\
1.14239732857811	-4.65990213095471 \\
1.20586384683245	-5.74340799256868 \\
1.26933036508679	-12.7243290500147 \\
1.33279688334112	-6.06770469873225 \\
1.39626340159546	-6.12834418375067 \\
1.4597299198498	-5.9445142410326 \\
1.52319643810414	-5.48488641132572 \\
1.58666295635848	-10.0976879703182 \\
1.65012947461282	-10.9271806841294 \\
1.71359599286716	-10.5015765029968 \\
1.7770625111215	-9.14721839866857 \\
1.84052902937584	-9.1427584616575 \\
1.90399554763018	-8.22429075990088 \\
1.96746206588452	-11.785353068347 \\
2.03092858413886	-7.15808057162908 \\
2.0943951023932	-13.2459036048378 \\
2.15786162064753	-12.5555164023598 \\
2.22132813890187	-9.82737134069593 \\
2.28479465715621	-11.1529512307817 \\
2.34826117541055	-15.5558411168884 \\
2.41172769366489	-10.6210851650249 \\
2.47519421191923	-13.1270693943339 \\
2.53866073017357	-13.8198523325338 \\
2.60212724842791	-13.0737257561727 \\
2.66559376668225	-10.024908800427 \\
2.72906028493659	-12.3902508813859 \\
2.79252680319093	-12.0457812880869 \\
2.85599332144527	-22.9621943128211 \\
2.91945983969961	-12.615849481223 \\
2.98292635795395	-11.4326590367134 \\
3.04639287620828	-10.5809466579845 \\
3.10985939446262	-10.2955099520882 \\
3.17332591271696	-10.5700674754946 \\
3.2367924309713	-8.4194147986417 \\
3.30025894922564	-12.5369353335509 \\
3.36372546747998	-12.0244130729592 \\
3.42719198573432	-9.73463103225246 \\
3.49065850398866	-13.2445617177569 \\
3.554125022243	-10.4596577377496 \\
3.61759154049734	-13.2529438161798 \\
3.68105805875168	-15.468126474705 \\
3.74452457700602	-9.84071453231885 \\
3.80799109526036	-19.7815281288075 \\
3.87145761351469	-13.7149079199217 \\
3.93492413176903	-13.5683207479987 \\
3.99839065002337	-16.9956976963886 \\
4.06185716827771	-12.6001203144949 \\
4.12532368653205	-16.3773501960904 \\
4.18879020478639	-14.6775315736444 \\
4.25225672304073	-8.7606725483799 \\
4.31572324129507	-10.8574533177887 \\
4.37918975954941	-11.2699732584048 \\
4.44265627780375	-14.2687764912224 \\
4.50612279605809	-16.5919848322137 \\
4.56958931431243	-15.2987044201685 \\
4.63305583256677	-11.2050452233429 \\
4.69652235082111	-13.8600275005054 \\
4.75998886907544	-7.59573107561616 \\
4.82345538732978	-11.3037771304612 \\
4.88692190558412	-9.14388167629639 \\
4.95038842383846	-10.0537731345592 \\
5.0138549420928	-15.4366882901524 \\
5.07732146034714	-10.1254803844769 \\
5.14078797860148	-7.38586407218227 \\
5.20425449685582	-13.8393073670075 \\
5.26772101511016	-6.91122348849911 \\
5.3311875333645	-8.36840750477432 \\
5.39465405161884	-13.1865940506337 \\
5.45812056987318	-9.50078436476079 \\
5.52158708812752	-13.0907590194137 \\
5.58505360638185	-13.4767561619825 \\
5.64852012463619	-9.13816269894019 \\
5.71198664289053	-7.80782286262393 \\
5.77545316114487	-8.97984439773249 \\
5.83891967939921	-14.9441875399564 \\
5.90238619765355	-8.19052341028716 \\
5.96585271590789	-7.21121634992635 \\
6.02931923416223	-15.1085315835341 \\
6.09278575241657	-9.39907871359652 \\
6.15625227067091	-7.16533933350971 \\
6.21971878892525	2.91189404314951 \\
6.28318530717959	12.5416583014942 \\
};

\addplot [dashed,black, mark = none, samples = 2, domain = 0:6.283]{-4.036402632837699};
\addplot [dashdotted,black, mark = none, samples = 2, domain = 0:6.283]{-24.0364026328377};

\end{axis}

\end{tikzpicture}
				
				\vspace{-20pt}
				
				\caption{ Result of $300$ Monte Carlo repetitions in the conditions specified 
						in Table~\ref{tab:physical_and_sim_parameters} evaluating the MSE in the
						estimation of $\slavephase$ by both PCP and LGS with respect to its actual
						value.  \label{fig:vs_phase_s_randomized}}
		\end{figure}
			
	\subsection{Average consistency} \label{sec:numres:average_consistency}
        
\begin{figure}
							\centering        
				\input{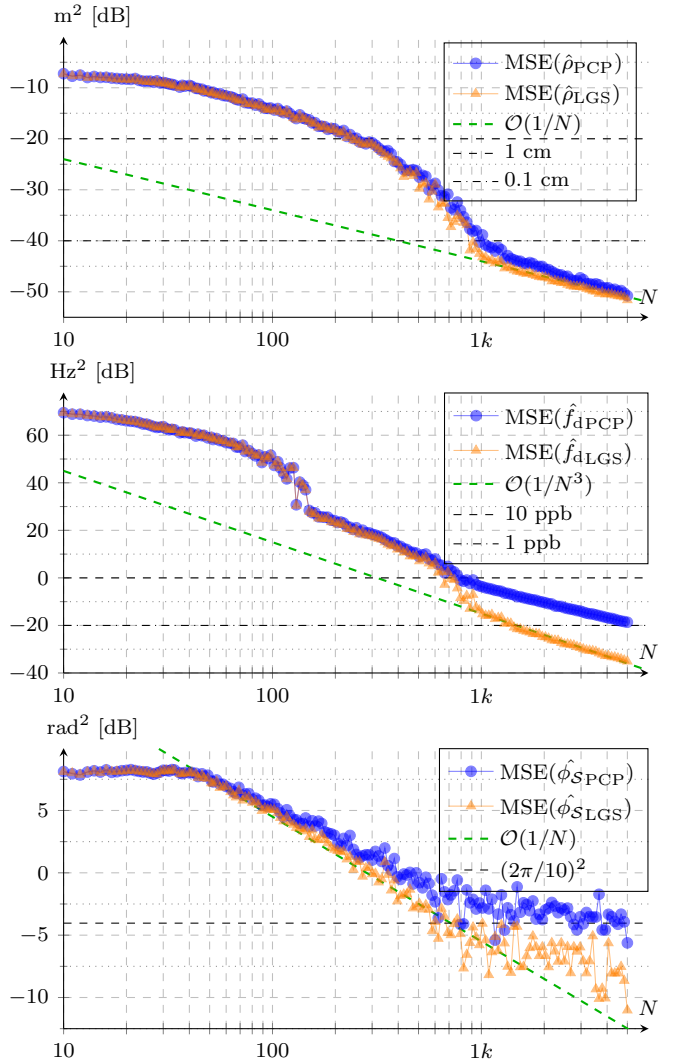}
				
				\vspace{-20pt}
				
				\caption{ Result of $2000$ Monte Carlo repetitions evaluating the MSE in the 
									estimation of $\rho$, $\freqd$ and $\slavephase$ by both PCP and LGS with
									respect to the sample size. The range and phase terms $\rho$ and 
									$\slavephase$ were randomized as specified in 
									Table~\ref{tab:physical_and_sim_parameters}, while
									$\freqd\sim\mathcal{U}([-200,-10)\cup[10,200))$, in order to avoid 
									the wildly irregular behavior of the estimators when
									$\freqd\in [-10,10)$, shown in Fig.~\ref{fig:vs_freq_d_randomized}.
									For reference and comparison, the convergence rates given by the 
									inverse Fisher information matrix we present in~\cite{AguilaPla2020} are portrayed
									by lines with the corresponding slope adjusted to fit the data.
									\label{fig:vs_sample_size_randomized} }
		\end{figure}

        An important property of estimators is consistency, i.e., the convergence
        of the MSE towards zero as the sample size increases.
		In order to characterize consistency beyond the example given 
		in~\cite{AguilaPla2020} with fixed physical parameters, in Fig.~\ref{fig:vs_sample_size_randomized}
		we report the average performance for randomized parameter values 
		for PCP and LGS in the estimation of the clock synchronization and 
		ranging parameters as a function of the sample size. However, we 
		sample frequency values in a reduced range, i.e., 
		$\freqd \in [-200,-10)\cup[10,200)$, in order to avoid the instability 
		of our estimators when $\freqd\in[-10,10)$. This provides an impression of 
		the estimators' performance within their range of probable use, and significantly
		reduces the amount of Monte Carlo repetitions necessary to obtain intelligible results.
		
		Fig.~\ref{fig:vs_sample_size_randomized} strengthens the conclusions from
		the analysis of the empirical results in~\cite{AguilaPla2020}. Indeed, 1) the results suggest that 
		both PCP and LGS are consistent for the estimation of $\rho$, $\freqd$ and $\slavephase$,
		2) LGS seems to have an asymptotically efficient rate of convergence with $N$ for 
		the estimation of both $\rho$ and $\freqd$, while the rate of convergence stagnates for 
		$\slavephase$. Here, we use as a reference the assymptotic rates of convergence for a linearized
		version of \eqref{eq:stoch_model}, which we derive from the Fisher information matrix
		in~\cite{AguilaPla2020}.
		
		Finally, from a practical point of view, average estimation errors under $0.1~\mathrm{cm}$ in
		the range parameter can be expected for $N\geq 1000$, and average estimation errors
		under $1~\mathrm{ppb}$ in the frequency parameter can be expected for $N\geq 1500$.
		For the phase parameter $\slavephase$, estimation errors below $2\pi/10$ can only be 
		expected with LGS and for $N\geq1000$.
		
	\subsection{Sensitivity to the inner and outer noises}
	
		\begin{figure}
				\centering
				\input{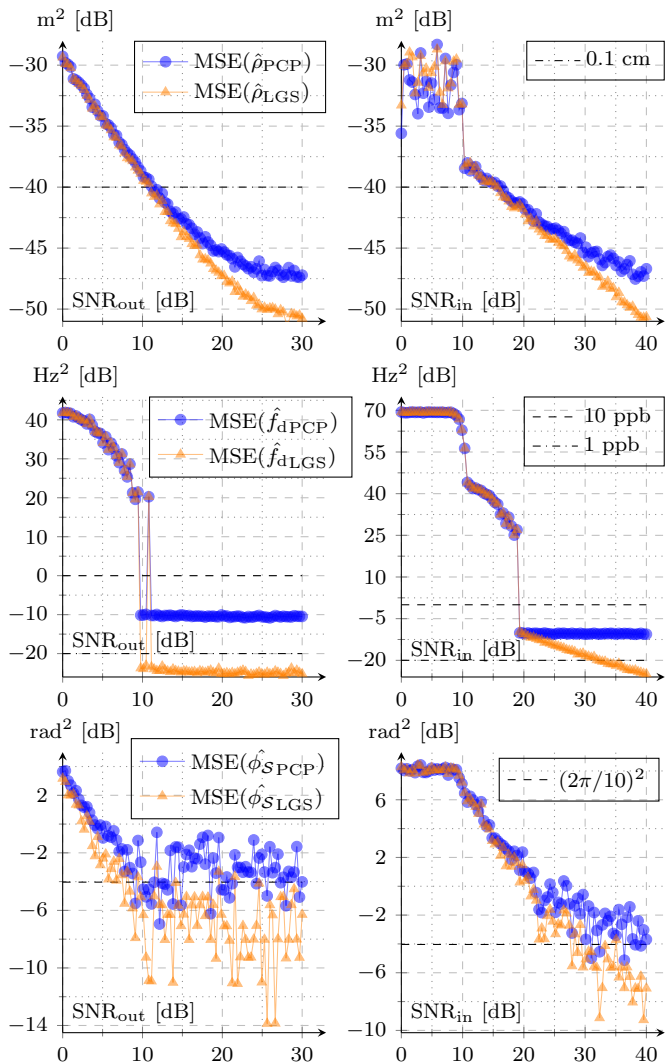}
				
				\vspace{-20pt}
				
				\caption{ Result of $1000$ Monte Carlo repetitions evaluating the MSE in the 
						estimation of the clock synchronization and ranging parameters $\rho$, 
						$\freqd$ and $\slavephase$, by both PCP and LGS with respect to the value of the SNRs. 
						Each SNR was fixed to its maximum when 
						the other was varied. The range and phase terms $\rho$ and 
									$\slavephase$ were randomized as specified in 
									Table~\ref{tab:physical_and_sim_parameters}, while
									$\freqd\sim\mathcal{U}([-200,-10)\cup[10,200))$, in order to avoid 
									the wildly irregular behavior of the estimators when
									$\freqd\in [-10,10)$, shown in Fig.~\ref{fig:vs_freq_d_randomized}.
						\label{fig:vs_SNRs_randomized}}
		\end{figure}
	
		All the results in \ref{fig:vs_freq_d_randomized} and 
		\ref{fig:vs_phase_s_randomized} (as well as those in~\cite{AguilaPla2020}) were obtained under the optimistic noise conditions 
		$\SNRin=40~\mathrm{dB}$ and $\SNRout=20~\mathrm{dB}$. In Fig.~\ref{fig:vs_SNRs_randomized}
		we report the average performance for randomized parameter values for PCP and LGS in the 
		estimation of the clock synchronization and ranging parameters as a function of either
		$\SNRin$ and $\SNRout$, when $\SNRout=30~\mathrm{dB}$ and $\SNRin=40~\mathrm{dB}$, respectively. 
		Similarly as in Section~\ref{sec:numres:average_consistency}, we sample $\freqd$ in the range
		$\freqd \in [-200,-10)\cup[10,200)$, in order to avoid the instability 
		of our estimators when $\freqd\in[-10,10)$.
		
		Fig.~\ref{fig:vs_SNRs_randomized} reveals that range estimation performance decays 
		progressively with the decrease of $\SNRout$ throughout the investigated range for both
		PCP and LGS. In contrast, the decrease of $\SNRin$ creates a progressive decay of performance
		only up to a breaking point around $\SNRin=10~\mathrm{dB}$. In this regime (very low $\SNRin$), 
		one could consider modeling the non-linear term in~\eqref{eq:model} as a uniform noise term and
		employ techniques tailored to the estimation of offsets in Gaussian plus uniform noise~\cite{Lundin2006}.
		Practically, both PCP and LGS achieve estimation accuracies below $0.1~\mathrm{cm}$
		only when $\SNRout\geq 10~\mathrm{dB}$ 
		for $\SNRin=40~\mathrm{dB}$, and only when $\SNRin\geq15~\mathrm{dB}$ for $\SNRout=30~\mathrm{dB}$.
		
		For frequency estimation, we see that both PCP and LGS are very robust to a decrease of $\SNRout$ up to
		a breaking point around $\SNRout = 10~\mathrm{dB}$, and LGS maintains an improvement of $15~\mathrm{dB}$
		over PCP for any $\SNRout$ above the breaking point. In contrast, when $\SNRin$ degrades, the breaking point
		for both PCP and LGS is $\SNRin=20~\mathrm{dB}$, and the improvement of LGS over PCP increases gradually as 
		$\SNRin$ increases, only reaching $15~\mathrm{dB}$ when $\SNRin=40~\mathrm{dB}$.
		Practically, on one hand, PCP achieves estimation accuracies below $10~\mathrm{ppb}$
		only when $\SNRout\geq 10~\mathrm{dB}$ 
		for $\SNRin=40~\mathrm{dB}$, and only when $\SNRin\geq20~\mathrm{dB}$ for $\SNRout=30~\mathrm{dB}$. 
		On the other hand, LGS consistently improves on it, reaching estimation accuracies below $1~\mathrm{ppb}$ 
		only when $\SNRout\geq 10~\mathrm{dB}$ 
		for $\SNRin=40~\mathrm{dB}$, and only when $\SNRin\geq30~\mathrm{dB}$ for $\SNRout=30~\mathrm{dB}$.
		Promising directions for improving frequency estimation could come from two different fronts. First,
		one could generalize the framework in~\cite{Lundin2006} to admit frequency estimation, aiming to 
		protect the resulting method for noises inside the nonlinearity beyond the breaking points in $\SNRin$.
		Second, one could use the structure of the sawtooth signal in techniques similar to~\cite{Camacho2008}
		to extract more information from the spectrum of the data.
		
		For phase estimation, our results are not conclusive, because the randomization of both the frequency and 
		phase parameters result in wide variability that would require further Monte Carlo averaging. Furthermore,
		while the unfavorable region of frequency parameters $\freqd\in[-10,10)$ has been avoided, the phase 
		parameters are still sampled from the whole range $\slavephase\in [0,2\pi)$, which includes the very challenging
		regions around the wrapping point (see Fig.~\ref{fig:vs_phase_s_randomized}). However,
		the results seem to suggest a progressive decay of performance for both PCP and LGS when either $\SNRout$ or
		$\SNRin$ degrade. Furthermore, we see that phase estimation is much more sensitive to a degradation of $\SNRin$
		than that of $\SNRout$. Possible improvements of phase estimation could be expected using LGS-type techniques
		complimented with better frequency parameter estimates.

\section{Conclusions} \label{sec:conc}

    In this paper, we provide practical and theoretical insights on the fundamental 
limitations of clock synchronization over networks in applications that require high 
ranging accuracy and low communication overhead. From a practical standpoint, we
show from first principles that using TDCs for measuring RTTs enables the use of a 
mathematical model that leads to very accurate ranging (e.g., accuracies beyond 
$0.1~\mathrm{cm}$ for $N\geq1000$ samples in Fig.~\ref{fig:vs_sample_size_randomized}) 
and remarkable frequency-synchronization performance (e.g., accuracies of $1~\mathrm{ppb}$
for $N\geq1500$ samples in the same figure), all with very simple estimation techniques.
Furthermore, we point at promising directions of research that could hold the key
to the improvement of these performance values, such as the extension 
of~\cite{Lundin2006} to improve range estimation when $\SNRin$ is very low (e.g., under 
$10~\mathrm{dB}$ in Fig.~\ref{fig:vs_SNRs_randomized}), or the 
use of techniques similar to~\cite{Camacho2008} to improve frequency synchronization.
In fact, in~\cite{AguilaPla2020} we provide a reference on the potential for improvement
in the form of approximated Cram\'{e}r-Rao lower bounds based on a linearization of the sawtooth model and standard results for Gaussian models.
Additionally, we pinpoint the weaknesses of both the model and our estimation 
strategies. First, we acknowledge that more research is needed to consistently obtain accurate 
phase synchronization (phase estimation accuracies beyond $2\pi/100$). Second, we identify 
that schemes that rely on this mathematical model are not best suited for absolute time 
synchronization because the offset delay only appears as part of a phase term, which leads to 
wrapping errors.

From a theoretical standpoint, we establish that RTT-based schemes for clock 
synchronization are characterized by similar fundamental limitations as those relying on 
time-stamped message exchanges, previously discovered by~\cite{Freris2010}. Namely,
synchronization is impossible with unknown path delays $\delayping$ and $\delaypong$ if 
one cannot assume some relation between $\delayping$, $\delaypong$ and the frequency
difference $\freqd$, e.g., that they are symmetric, i.e., $\delayping=\delaypong$.
Furthermore, we have discovered a surprising property of sawtooth signal models, i.e., 
that adding noise inside the non-linearity allows for the joint identifiability 
of its offset and phase terms (cf. Lemma~\ref{lem:unid-nojit} and Theorem~\ref{th:id}). 
This result challenges the convention of random variations as a negative component of a model,
and is of use beyond our application domain.

\appendices
    
    \section{Proof of identifiability} \label{app:id_proof}
    
        \begin{IEEEproof}[Proof of Theorem~\ref{th:id}, Identifiability of \eqref{eq:model} when $\sigma_v>0$]
    The backward implication of identifiability, i.e., that the same parameters
    lead to the same data distribution (see Def.~\ref{def:id}), is immediately clear
    from \eqref{eq:model}.
    
    In the following, we will show that under the conditions above, the forward
    implication is also true, i.e., that 
    \begin{IEEEeqnarray}{c}\label{eq:equaldistr}
      \pdf{\datarv}{\parsone}{\datasa} - \pdf{\datarv}{\parstwo}{\datasa}=0,
      \forall \datasa \in \reals^N
    \end{IEEEeqnarray}
    implies that $\parsone = \parstwo$, where $\parsone$ and $\parstwo$ are two vectors of parameters, and
    we denote their respective components by the same super-index, i.e., 
    $\alpha^{(i)}$, $\psi^{(i)}$, $\beta^{(i)}$, and $\gamma^{(i)}$ for $i\in\lbrace1,2\rbrace$.
    Consequently, we assume \eqref{eq:equaldistr}, and will reach the conclusion
    that $\parsone = \parstwo$.
    We start by defining the random process 
    \begin{IEEEeqnarray}{rl}
      P[n]\, &= \alpha +  \psi \modu\left(\beta n + \gamma + V[n] \right) \label{eq:P}\,,
    \end{IEEEeqnarray}
    and observe that \eqref{eq:model} implies that $Y[n]=W[n]+P[n]$.
    Because $W[n]$ and $V[n]$ are assumed independent (see \eqref{eq:model} and the discussion in Section~\ref{sec:SysMod}) and 
    $W[n]\sim\normal{0}{\sigma_w^2}$, we have that, if $*$ denotes a convolution and $\varphi(\cdot)$ is the PDF of 
    the standard normal distribution $\normal{0}{1}$, the distribution of $Y[n]$ is, for any given $n\in\listintzero{N-1}$,
    \begin{IEEEeqnarray}{c}
      \label{eq:ConvProp}
      \pdf{Y}{\pars}{y} = \left[ \varphi\!\left(\frac{\tau}{\sigma_w}\right) * \pdf{P}{\pars}{\tau}\right](y)\,.
    \end{IEEEeqnarray}
    Then, consider the difference between the PDFs of $Y[n]$ that correspond
    to the parameter vectors $\parsone$ and $\parstwo$, i.e.,
    \begin{IEEEeqnarray*}{rl}
      \Delta_Y(y)\, & = \pdf{Y}{\parsone}{y} - \pdf{Y}{\parstwo}{y} \\
      &= \left[ 
	\varphi\!\left(\frac{\tau}{\sigma_w}\right) * 
	\underbrace{ \left( \pdf{P}{\parsone}{\tau} - \pdf{P}{\parstwo}{\tau} \right)}_{\Delta_P(\tau)} \right](y)\,,
    \end{IEEEeqnarray*}
    and observe that \eqref{eq:equaldistr} implies that
    $\Delta_Y(y)=0$, $\forall y\in\reals$ and $\forall n \in \listintzero{N-1}$.
    Because $\varphi(\tau)>0$ for any $\tau\in\reals$, we have that $\Delta_Y(y)=0$, 
    $\forall y\in\reals$ if and only if $\Delta_P(p) = 0$, $\forall p\in\reals$. 
    Because $P[n]\in[\alpha,\alpha+\psi)$ if $\psi>0$ and 
    $P[n]\in(\alpha+\psi,\alpha]$ if $\psi<0$, this can only happen if 
    $\alpha^{(1)} = \alpha^{(2)}$ and $\psi^{(1)}=\psi^{(2)}$, i.e.,
    \begin{IEEEeqnarray}{c} \label{eq:rhos}
      \!\!\Delta_Y(y)=0, \forall y\in\reals \Rightarrow \alpha^{(1)} = \alpha^{(2)}\mbox{ and }\psi^{(1)}=\psi^{(2)}\,.
    \end{IEEEeqnarray}
    Therefore, the ambiguity between $\alpha$ and $\gamma$ that led to Lem.~\ref{lem:unid-nojit} has been resolved by the 
    presence of the noise term $V[n]$ inside the non-linearity. This effect is exemplified by the simple example of Fig.~\ref{fig:ide_jitt}, 
    which studies the same exact situation as Fig.~\ref{fig:unidentifiable}b), but with $\sigma_v>0$.
    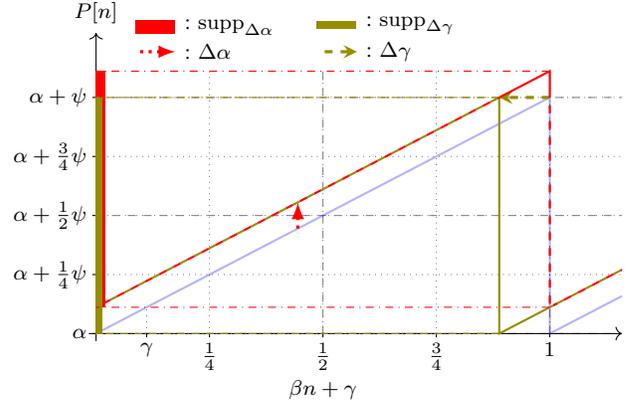
\begin{figure}
			\centering
			\def\xs{.96} \def\ys{.5} \def\fs{\footnotesize}
\def\pini{0.7}
\def\pshift{0.7}
\def\xfinal{2*pi+1}
\pgfmathsetmacro\bythree{1/3}
\pgfmathsetmacro\byxsc{1/\xs}
\begin{tikzpicture}[xscale=\xs,yscale=\ys]
    \node[anchor=west] (RhoName) at (pi*0.35,7.5) {\fs: $\Delta\alpha$};
    \draw [latex-,color=red, very thick, dotted] (RhoName.west) --++ (-0.5,0);
    \node[anchor=south west,yshift=3pt] (SuppRhoName) at (RhoName.west) {\fs: $\mathrm{supp}_{\Delta\alpha}$ };
    \node[anchor=east,rectangle,fill = red, fill opacity=1, minimum width=15pt, minimum height=5pt, inner sep=0pt] (RhoSign) at (SuppRhoName.west) {};
    \node[anchor=west] (PhaseName) at (pi*1.15,7.5) {\fs: $\Delta\gamma$};
    \draw [stealth-,color=olive, very thick, dashed] (PhaseName.west) --++ (-0.5,0);
    \node[anchor=south west,yshift=3pt] (SuppPhaseName) at (PhaseName.west) {\fs: $\mathrm{supp}_{\Delta\gamma}$ };
    \node[anchor=east,rectangle,fill = olive, fill opacity=1, minimum width=15pt, minimum height=3pt, inner sep=0pt] (PhaseSign) at (SuppPhaseName.west) {};
    

        \draw [dotted,xstep=.5*pi,ystep=.5*pi, very thin] (0,0) grid (\xfinal,7);
        \draw [dashed,xstep=pi,ystep=pi, gray, very thin] (0,0) grid (\xfinal,7);
        \node at (\pini,0) [anchor=north] {\fs  $\gamma$};
        \draw (\pini,-0.1) -- (\pini,0);
        \draw [dotted,very thin] (\pini,0) -- (\pini,\pini);
        \node (EndX) at (\xfinal,0) [anchor=west] {};
        \node at (pi,-1) [anchor=north] {\fs $\beta n + \gamma$};
        \node (EndY) at (0,8) [anchor=south] {\fs $P[n]$};
        \draw[->] (-0.1,0) -- (EndX);
        \draw[->] (0,-0.2) -- (EndY);
        \foreach \x/\xl in {.5*pi/\tfrac{1}{4}, pi/\tfrac{1}{2},   0.5*3*pi/\tfrac{3}{4}, 2*pi/1 } 
        {
	        \draw (\x,0) node[anchor=north] {\fs $\xl$};
	        \draw (\x,-0.1) -- (\x,0);
	    }
        \foreach \y/\yl in {0/\alpha, .5*pi/\alpha+\tfrac{1}{4}\psi, pi/\alpha+\tfrac{1}{2}\psi,   0.5*3*pi/\alpha+\tfrac{3}{4}\psi, 2*pi/\alpha+\psi}
        {
            \draw (0,\y) node[anchor=east] {\fs $\yl$};
            \draw (-.1*.5*\byxsc,\y) -- (0,\y);
        }
			\draw[color=blue, thick, opacity=0.3] (0,0) -- (2*pi,2*pi) -- (2*pi,0) -- (2*pi+0.1, 0.1);
			\draw[color=blue, thick, opacity=0.3] (2*pi+0.1, 0.1) -- (\xfinal,\xfinal-2*pi);		
		
			\foreach \p in { 2*pi}
			{
				\draw[-stealth,color=olive, very thick, dashed] (\p,\p) -- (\p-\pshift,\p);
			}

			\foreach \p in { 2*pi*\bythree+\pini}
			{
				\draw[-latex,color=red, very thick, dotted] (\p,\p) -- (\p,\p+\pshift);
			}

		\draw[dashed,color=olive,thick] (0,\pshift) -- (2*pi-\pshift,2*pi);
		\draw[color=olive,thick] (2*pi-\pshift,2*pi)-- (2*pi-\pshift, 0) -- (2*pi+0.1,0.1+\pshift);
		\draw[color=olive,dashed,thick] (2*pi+0.1,0.1+\pshift) -- (\xfinal,\xfinal-2*pi+\pshift);
		 
		\draw[dashed,color=red, thick,shorten <=3] (0,\pshift) -- (2*pi-\pshift,2*pi);
		\draw[color=red, thick](2*pi-\pshift,2*pi)--(2*pi,2*pi+\pshift)-- (2*pi,2*pi);
		\draw[dashed,color=red, thick,shorten <=3] (2*pi,2*pi) -- (2*pi,\pshift) -- (2*pi+0.1, 0.1+\pshift);
		\draw[dashed,color=red, thick,shorten <=3] (2*pi+0.1, 0.1+\pshift) -- (\xfinal,\xfinal-2*pi+\pshift);

		\fill[opacity=1, color=red] (-0.01,\pshift) rectangle (0.12,2*pi+\pshift);
		\fill[opacity=1, color=olive] (-0.01,0) rectangle (0.08,2*pi);
		
		\draw[dashdotted,color=red] (2*pi,2*pi+\pshift) -- (0,2*pi+\pshift);
		\draw[dashdotted,color=red] (2*pi, \pshift) -- (0,\pshift);
		
		\draw[densely dashdotted,color=olive] (2*pi-\pshift,2*pi) -- (0,2*pi);
		\draw[densely dashdotted,color=olive] (2*pi-\pshift, 0) -- (0,0);
\end{tikzpicture}
			
			\vspace{-10pt}
			
			\caption{Example on how the same changes of parameters $\Delta \alpha \geq 0$ and $\Delta \gamma \geq 0$
				that exemplified in Fig.~\ref{fig:unidentifiable}(b) that the model $\model$ with $\sigma_v=0$ was not
				identifiable yield different supports of $P[n]$, i.e., 
				$\mathrm{supp}_{\Delta \alpha}$ and $\mathrm{supp}_{\Delta\gamma}$, when $\sigma_v>0$.
				Accordingly, changes in $\alpha$ and $\gamma$, however small, will lead to different distributions of $Y[n]$.
				\label{fig:ide_jitt}
			}    
		\end{figure}    
    
    Consider then the random process $Q[n]$ such that 
    $Q[n] = \mod{2\pi}\left(2\pi \beta n + 2\pi\gamma + 2\pi V[n] \right)$, and observe that 
    \begin{IEEEeqnarray*}{c}
      P[n] = \alpha + \frac{\psi}{2\pi} Q[n]\,.
    \end{IEEEeqnarray*}
    $Q[n]$ has a wrapped-normal distribution~\cite[p.~50]{Mardia2009} that is monomodal with mode 
    $\mod{2\pi}\left(2\pi \beta n + \gamma \right)$
    and therefore, $P[n]$ is monomodal with mode 
    $\alpha+ \psi \modu\left(\beta n + \gamma \right)$. 
    Because $\Delta_Y(y)=0$, $\forall y \in \reals$ implies that $\Delta_P(p) = 0$, $\forall p\in\reals$, we have that
    in particular, the mode of $P[n]$ under $\parsone$ and $\parstwo$ must also be the same, which, because 
    $\alpha^{(1)} = \alpha^{(2)}$ and $\psi^{(1)}=\psi^{(2)}$, implies that
    \begin{IEEEeqnarray}{c}\label{eq:condfin}
      \modu\left( \beta^{(1)} n + \gamma^{(1)} \right) =
      \modu\left( \beta^{(2)} n + \gamma^{(2)} \right)\,.
    \end{IEEEeqnarray}
    Because $N\geq 2$, we know that \eqref{eq:condfin} must be verified at least for $n\in\lbrace0,1\rbrace$.
    For $n=0$, because $\gamma^{(1)},\gamma^{(2)}\in[0,1)$, \eqref{eq:condfin} implies that
    \begin{IEEEeqnarray*}{c}
		\modu\left( \gamma^{(1)} \right) = \modu\left( \gamma^{(2)} \right) 
		\Rightarrow \gamma^{(1)} = \gamma^{(2)}\,.
    \end{IEEEeqnarray*}
    For $n=1$, because $\gamma^{(1)} = \gamma^{(2)}$, \eqref{eq:condfin} implies that there are $K_1,K_2 \in \integ$ with $Q = K_2 - K_1$ such that
    \begin{IEEEeqnarray*}{c}
		\beta^{(1)} + K_1 = \beta^{(2)} + K_2 \Rightarrow \beta^{(1)} 
		= Q + \beta^{(2)}\,.
    \end{IEEEeqnarray*}
    Because $\beta^{(1)},\beta^{(2)} \in \left[-\frac{1}{2},\frac{1}{2}\right)$
    this can only be fulfilled for $Q=0$, and thus, $\beta^{(1)} = \beta^{(2)}$. In conclusion, \eqref{eq:equaldistr}
    implies that $\alpha^{(1)} = \alpha^{(2)}$, $\psi^{(1)}=\psi^{(2)}$,  $\gamma^{(1)} = \gamma^{(2)}$, and $\beta^{(1)} = \beta^{(2)}$, i.e., $\parsone = \parstwo$.

  \end{IEEEproof}

\section*{Acknowledgments}

  The authors thank Dr. Hugo Tullberg for referring us to the existing results on
  circular statistics, which were fundamental to finalize the proof of Theorem~\ref{th:id}. 
  The authors thank Gerard Farr\'{e} for lengthy discussions on dynamical systems, the 
  $\mod{}$ equivalence, and first directing us to the concept of the orbit of a rotation of 
  the circle.



\bibliographystyle{IEEEtranMod}
\bibliography{IEEEabrv,clock-sync}

\begin{thebibliography}{10}
\providecommand{\url}[1]{#1}
\csname url@samestyle\endcsname
\providecommand{\newblock}{\relax}
\providecommand{\bibinfo}[2]{#2}
\providecommand{\BIBentrySTDinterwordspacing}{\spaceskip=0pt\relax}
\providecommand{\BIBentryALTinterwordstretchfactor}{4}
\providecommand{\BIBentryALTinterwordspacing}{\spaceskip=\fontdimen2\font plus
\BIBentryALTinterwordstretchfactor\fontdimen3\font minus
  \fontdimen4\font\relax}
\providecommand{\BIBforeignlanguage}[2]{{%
\expandafter\ifx\csname l@#1\endcsname\relax
\typeout{** WARNING: IEEEtran.bst: No hyphenation pattern has been}%
\typeout{** loaded for the language `#1'. Using the pattern for}%
\typeout{** the default language instead.}%
\else
\language=\csname l@#1\endcsname
\fi
#2}}
\providecommand{\BIBdecl}{\relax}
\BIBdecl

\bibitem{Lamport1978}
L.~Lamport, ``Time, clocks, and the ordering of events in a distributed
  system,'' \emph{Communications of the Association for Computing Machinery},
  vol.~21, no.~7, pp. 558--565, Jul. 1978.

\bibitem{Freris2010}
N.~M. Freris, H.~Kowshik, and P.~R. Kumar, ``Fundamentals of large sensor
  networks: {C}onnectivity, capacity, clocks, and computation,''
  \emph{Proceedings of the IEEE}, vol.~98, no.~11, pp. 1828--1846, Nov. 2010.

\bibitem{Etzlinger2014}
B.~Etzlinger, H.~Wymeersch, and A.~Springer, ``Cooperative synchronization in
  wireless networks,'' \emph{IEEE Transactions on Signal Processing}, vol.~62,
  no.~11, pp. 2837--2849, Jun. 2014.

\bibitem{Xia2018}
W.~Xia and M.~Cao, ``Determination of clock synchronization errors in
  distributed networks,'' \emph{SIAM Journal on Control and Optimization},
  vol.~56, no.~2, pp. 610--632, 2018.

\bibitem{Geng2018}
Y.~Geng, S.~Liu, Z.~Yin, A.~Naik, B.~Prabhakar, M.~Rosunblum, and A.~Vahdat,
  ``Exploiting a natural network effect for scalable, fine-grained clock
  synchronization,'' in \emph{Proceedings of the 15th USENIX Conference on
  Networked Systems Design and Implementation}, ser. NSDI'18.\hskip 1em plus
  0.5em minus 0.4em\relax Berkeley, CA, USA: USENIX Association, 2018, pp.
  81--94.

\bibitem{Etzlinger2018}
B.~Etzlinger and H.~Wymeersch, ``Synchronization and localization in wireless
  networks,'' \emph{Foundations and Trends{\textregistered} in Signal
  Processing}, vol.~12, no.~1, pp. 1--106, 2018.

\bibitem{Zachariah2017}
D.~Zachariah, S.~Dwivedi, P.~H\"{a}ndel, and P.~Stoica, ``Scalable and passive
  wireless network clock synchronization in {LoS} environments,'' \emph{IEEE
  Transactions on Wireless Communications}, vol.~16, no.~6, pp. 3536--3546,
  Jun. 2017.

\bibitem{Koivisto2017}
M.~Koivisto, M.~Costa, J.~Werner, K.~Heiska, J.~Talvitie, K.~Leppänen,
  V.~Koivunen, and M.~Valkama, ``Joint device positioning and clock
  synchronization in {5G} ultra-dense networks,'' \emph{IEEE Transactions on
  Wireless Communications}, vol.~16, no.~5, pp. 2866--2881, May 2017.

\bibitem{Event2019}
{Event Horizon Telescope Collaboration et al.}, ``First {M87} event horizon
  telescope results. {I}. {T}he shadow of the supermassive black hole,''
  \emph{Astrophysical Journal Letters}, vol. 875, no.~1, p.~L1, 2019.

\bibitem{Chen2017}
P.~Chen and A.~Babakhani, ``3-{D} radar imaging based on a synthetic array of
  30-{GHz} impulse radiators with on-chip antennas in 130-nm {SiGe} {BiCMOS},''
  \emph{IEEE Transactions on Microwave Theory and Techniques}, vol.~65, no.~11,
  pp. 4373--4384, Nov. 2017.

\bibitem{Jamali2018}
B.~Jamali and A.~Babakhani, ``A self-mixing picosecond impulse receiver with an
  on-chip antenna for high-speed wireless clock synchronization,'' \emph{IEEE
  Transactions on Microwave Theory and Techniques}, vol.~66, no.~5, pp.
  2313--2324, May 2018.

\bibitem{He2014}
J.~He, P.~Cheng, L.~Shi, J.~Chen, and Y.~Sun, ``Time synchronization in {WSN}s:
  a maximum-value-based consensus approach,'' \emph{IEEE Transactions on
  Automatic Control}, vol.~59, no.~3, pp. 660--675, Mar. 2014.

\bibitem{Carli2014}
R.~Carli and S.~Zampieri, ``Network clock synchronization based on the
  second-order linear consensus algorithm,'' \emph{IEEE Transactions on
  Automatic Control}, vol.~59, no.~2, pp. 409--422, Feb. 2014.

\bibitem{Kim2017}
K.~S. Kim, S.~Lee, and E.~G. Lim, ``Energy-efficient time synchronization based
  on asynchronous source clock frequency recovery and reverse two-way message
  exchanges in wireless sensor networks,'' \emph{IEEE Transactions on
  Communications}, vol.~65, no.~1, pp. 347--359, Jan. 2017.

\bibitem{Bolognani2016}
S.~Bolognani, R.~Carli, E.~Lovisari, and S.~Zampieri, ``A randomized linear
  algorithm for clock synchronization in multi-agent systems,'' \emph{IEEE
  Transactions on Automatic Control}, vol.~61, no.~7, pp. 1711--1726, Jul.
  2016.

\bibitem{Freris2011}
N.~M. Freris, S.~R. Graham, and P.~R. Kumar, ``Fundamental limits on
  synchronizing clocks over networks,'' \emph{IEEE Transactions on Automatic
  Control}, vol.~56, no.~6, pp. 1352--1364, Jun. 2011.

\bibitem{Zheng2010}
J.~Zheng and Y.-C. Wu, ``Joint time synchronization and localization of an
  unknown node in wireless sensor networks,'' \emph{IEEE Transactions on Signal
  Processing}, vol.~58, no.~3, pp. 1309--1320, Mar. 2010.

\bibitem{Chepuri2013}
S.~P. Chepuri, R.~T. Rajan, G.~Leus, and A.-J. van~der Veen, ``Joint clock
  synchronization and ranging: {A}symmetrical time-stamping and passive
  listening,'' \emph{IEEE Signal Processing Letters}, vol.~20, no.~1, pp.
  51--54, Jan. 2013.

\bibitem{Dwivedi2015}
S.~Dwivedi, A.~D. Angelis, D.~Zachariah, and P.~H\"{a}ndel, ``Joint ranging and
  clock parameter estimation by wireless round trip time measurements,''
  \emph{IEEE Journal on Selected Areas in Communications}, vol.~33, no.~11, pp.
  2379--2390, Nov. 2015.

\bibitem{Gholami2015}
M.~R. Gholami, S.~Dwivedi, M.~Jansson, and P.~Händel, ``Ranging without time
  stamps exchanging,'' in \emph{2015 IEEE International Conference on
  Acoustics, Speech and Signal Processing (ICASSP)}, Apr. 2015, pp. 3981--3985.

\bibitem{Angelis2013}
A.~D. Angelis, S.~Dwivedi, and P.~H\"{a}ndel, ``Characterization of a flexible
  {UWB} sensor for indoor localization,'' \emph{IEEE Transactions on
  Instrumentation and Measurement}, vol.~62, no.~5, pp. 905--913, May 2013.

\bibitem{Patwari2005}
N.~Patwari, J.~N. Ash, S.~Kyperountas, A.~O. Hero, R.~L. Moses, and N.~S.
  Correal, ``Locating the nodes: {C}ooperative localization in wireless sensor
  networks,'' \emph{IEEE Signal Processing Magazine}, vol.~22, no.~4, pp.
  54--69, Jul. 2005.

\bibitem{Nilsson2014}
J.-O. Nilsson, J.~Rantakokko, P.~Händel, I.~Skog, M.~Ohlsson, and K.~V.~S.
  Hari, ``Accurate indoor positioning of firefighters using dual foot-mounted
  inertial sensors and inter-agent ranging,'' in \emph{2014 IEEE/ION Position,
  Location and Navigation Symposium (PLANS 2014)}, May 2014, pp. 631--636.

\bibitem{Ciftci2001}
M.~Ciftci and D.~B. Williams, ``Optimal estimation and sequential channel
  equalization algorithms for chaotic communications systems,'' \emph{EURASIP
  J. Appl. Signal Process.}, vol. 2001, no.~1, pp. 249--256, Jan. 2001.

\bibitem{Drake2007}
D.~F. Drake and D.~B. Williams, ``Linear, random representations of chaos,''
  \emph{IEEE Transactions on Signal Processing}, vol.~55, no.~4, pp.
  1379--1389, Apr. 2007.

\bibitem{AguilaPla2020}
P.~del Aguila~Pla, L.~Pellaco, S.~Dwivedi, P.~H\"{a}ndel, and J.~Jald\'{e}n,
  ``Clock synchronization over networks using sawtooth models,'' in \emph{2020
  IEEE International Conference on Acoustics, Speech and Signal Processing
  (ICASSP)}, 2020.

\bibitem{GITHUB}
P.~del Aguila~Pla and L.~Pellaco, ``{c}lock sync and range,'' GitHub
  repository, \url{https://github.com/poldap/clock_sync_and_range}, 2018.

\bibitem{Walter1997}
E.~Walter and L.~Pronzato, \emph{Identification of parametric models from
  experimental data}.\hskip 1em plus 0.5em minus 0.4em\relax Springer-Verlag,
  1997, ch. 2.6.1 Identifiability, pp. 20--32.

\bibitem{Casella2002}
G.~Casella and R.~L. Berger, \emph{Statistical inference}, 2nd~ed.\hskip 1em
  plus 0.5em minus 0.4em\relax Duxbury Pacific Grove, CA, 2002.

\bibitem{Katok1995}
A.~Katok and B.~Hasselblatt, \emph{Introduction to the modern theory of
  dynamical systems}.\hskip 1em plus 0.5em minus 0.4em\relax Cambridge
  University Press, 1995.

\bibitem{Vaart1998}
A.~W. van~der Vaart, \emph{Asymptotic statistics}.\hskip 1em plus 0.5em minus
  0.4em\relax Cambridge University Press, 1998.

\bibitem{Mardia2009}
K.~V. Mardia and P.~E. Jupp, \emph{Directional statistics}.\hskip 1em plus
  0.5em minus 0.4em\relax John Wiley \& Sons, 2009, vol. 494.

\bibitem{Noh2007}
K.-L. Noh, Q.~M. Chaudhari, E.~Serpedin, and B.~W. Suter, ``Novel clock phase
  offset and skew estimation using two-way timing message exchanges for
  wireless sensor networks,'' \emph{IEEE Transactions on Communications},
  vol.~55, no.~4, pp. 766--777, Apr. 2007.

\bibitem{Skog2010}
I.~Skog and P.~H\"{a}ndel, ``Synchronization by two-way message exchanges:
  {C}ramér-{R}ao bounds, approximate maximum likelihood, and offshore
  submarine positioning,'' \emph{IEEE Transactions on Signal Processing},
  vol.~58, no.~4, pp. 2351--2362, Apr. 2010.

\bibitem{Kay2000}
S.~Kay, ``Can detectability be improved by adding noise?'' \emph{IEEE Signal
  Processing Letters}, vol.~7, no.~1, pp. 8--10, Jan. 2000.

\bibitem{Chen2007}
H.~Chen, P.~K. Varshney, S.~Kay, and J.~H. Michels, ``Theory of the stochastic
  resonance effect in signal detection: {P}art {I}---{F}ixed detectors,''
  \emph{IEEE Transactions on Signal Processing}, vol.~55, no.~7, pp.
  3172--3184, Jul. 2007.

\bibitem{Chen2008}
H.~Chen and P.~K. Varshney, ``Theory of the stochastic resonance effect in
  signal detection---{P}art {II}: {V}ariable detectors,'' \emph{IEEE
  Transactions on Signal Processing}, vol.~56, no.~10, pp. 5031--5041, Oct.
  2008.

\bibitem{Kay2008}
S.~Kay, ``Noise enhanced detection as a special case of randomization,''
  \emph{IEEE Signal Processing Letters}, vol.~15, pp. 709--712, 2008.

\bibitem{Chen2014}
H.~Chen, L.~R. Varshney, and P.~K. Varshney, ``Noise-enhanced information
  systems,'' \emph{Proceedings of the IEEE}, vol. 102, no.~10, pp. 1607--1621,
  Oct. 2014.

\bibitem{Schuchman1964}
L.~Schuchman, ``Dither signals and their effect on quantization noise,''
  \emph{IEEE Transactions on Communication Technology}, vol.~12, no.~4, pp.
  162--165, Dec. 1964.

\bibitem{Lundin2006}
H.~Lundin, M.~Skoglund, and P.~H\"{a}ndel, ``On the estimation of quantizer
  reconstruction levels,'' \emph{IEEE Transactions on Instrumentation and
  Measurement}, vol.~55, no.~6, pp. 2176--2182, Dec. 2006.

\bibitem{Camacho2008}
A.~Camacho and J.~G. Harris, ``A sawtooth waveform inspired pitch estimator for
  speech and music,'' \emph{The Journal of the Acoustical Society of America},
  vol. 124, no.~3, pp. 1638--1652, 2008.

\end{thebibliography}

\enlargethispage{-38pt}

\begin{IEEEbiography}[{\includegraphics[width=1in,height=1.25in,clip,keepaspectratio]{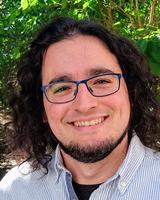}}]{Pol del Aguila Pla}
    (S'15-M'19) received a double degree in telecommunications and electrical engineering from the Universitat Polit\`{e}cnica de Catalunya 
    (UPC) and the KTH Royal Institute of Technology in 2014, and a Ph.D. in electrical engineering from the KTH Royal Institute of Technology in 2019. During his Ph.D., he investigated several inverse problems in signal processing applications such as biomedical imaging and clock synchronization over networks. Since October 2019, Pol is a research staff scientist at the Center for Biomedical Imaging
    (CIBM) in Switzerland, and a postdoctoral researcher at the EPFL's Biomedical Imaging Group in Lausanne,
    Switzerland.
    
    Pol is a reviewer for the \textsc{IEEE Transactions on Signal Processing}, the \textsc{IEEE Open Journal of Signal Processing}, the \textsc{IEEE Wireless Communications Letters}, and \textsc{Elsevier Signal Processing}, and he is part of the local committee of \textsc{ICASSP 2020}, as well as a reviewer
    for \textsc{ICML 2020}.
\end{IEEEbiography}


\begin{IEEEbiography}[{\includegraphics[height=1.25in,width=1in,clip,keepaspectratio]{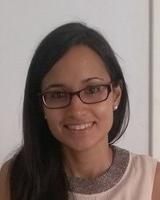}}]{Lissy Pellaco}
    (S'18) received \emph{summa cum laude} the M.Sc. degree in Multimedia Signal Processing and Telecommunication Networks from the University of Genoa, Italy, in 2016. After an international assignment on Industrial Automation Networks at ABB in Cleveland, Ohio, she is currently enrolled in a Ph.D. program in Machine Learning applied to Radio Networks at KTH, under the supervision of Joakim Jaldén. She is also affiliated to the Wallenberg AI, Autonomous Systems and Software Program (WASP).
    In her Ph.D. program, Lissy is investigating the merger of traditional parameterized and machine learning-based approaches to wireless communication, both satellite and terrestrial. She is also a
    reviewer for EUSIPCO 2019, and, since February 2019, she is part of the coordination group of the Female PhD Student Network at the School of Electrical Engineering and Computer Science in KTH.

    In December 2018 she received the graduate award issued by the Italian Association of Electrical, Electronics, Automation, Information and Communication Technology (AEIT) to the most promising graduate in ICT Engineering. In April 2018 she was awarded the Excellence grant by the Executive Committee of the Doctoral Program council of the school of Electrical Engineering and Computer Science at KTH. Also, in 2013 she was the recipient of the “ASING” and “IRIS INGEGNERIA” awards issued by the University of Genoa to the most brilliant first-year B.Sc. students.    
\end{IEEEbiography}


\begin{IEEEbiography}[{\includegraphics[height=1.25in,width=1in,clip,keepaspectratio]{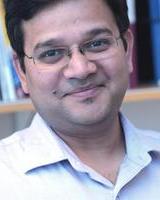}}]{Satyam Dwivedi}
    (M'14) received his MS and PhD degrees from the Indian Institute of Science, Bangalore, India. He is a senior researcher with Ericsson Research in Stockholm. He has been a researcher and a teacher at KTH, Stockholm. His research interests include wireless positioning, time synchronization, wireless propagation, and wireless testbeds.
\end{IEEEbiography}


\begin{IEEEbiography}[{\includegraphics[width=1in,height=1.25in,clip,keepaspectratio]{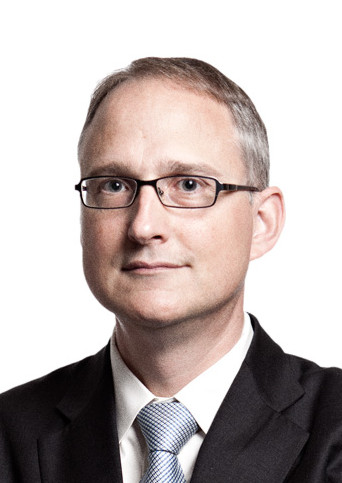}}]{Peter H\"{a}ndel}
    Peter Händel (S’88–M’94–SM’98) received the
    M.Sc. degree in engineering physics and the
    Lic.Eng. and Ph.D. degrees in automatic control
    from the Department of Technology, Uppsala University, 
    Uppsala, Sweden, in 1987, 1991, and 1993,
    respectively. He held a part-time position as an Associate
    Director of research with the Swedish Defense
    Research Agency from 2000 to 2006. In 2010, he
    joined the Indian Institute of Science, Bangalore,
    India, as a Guest Professor. He was a Guest Professor
    with the University of Gävle between 2007 and
    2013. Since 1997, he has been with the KTH Royal Institute of Technology,
    Stockholm, Sweden, where he is currently a Professor of signal processing
    with the School of Electrical Engineering and Computer Science. He has
    authored over 300 scientific publications. He was a recipient of a number of
    awards, including the \textsc{IEEE Transactions on Intelligent Transportation Systems}
    Best Survey Paper Award. He is the former President of the IEEE Finland Joint Signal Processing
    and Circuits and Systems Chapter and the former President of the IEEE
    Sweden Signal Processing Chapter. He was an Associate Editor of the 
    \textsc{IEEE Transactions on Signal Processing}.

\end{IEEEbiography}

\begin{IEEEbiography}[{\includegraphics[width=1in,height=1.25in,clip,keepaspectratio]{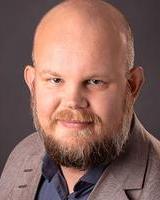}}]{Joakim Jald\'{e}n} 
    (S’03-M’08-S’13) received the M.Sc.\ and Ph.D.\ in electrical engineering from the KTH Royal Institute of Technology, Stockholm, Sweden in 2002 and 2007 respectively. From July 2007 to June 2009 he held a post-doctoral research position at the Vienna University of Technology, Vienna, Austria. He also studied at Stanford University, CA, USA, from September 2000 to May 2002, and worked at ETH, Z\"{u}rich, Switzerland, as a visiting researcher, from August to September, 2008. In July 2009 he returned to KTH, where he is now a professor of signal processing. He was an associate editor for the \textsc{IEEE Communications Letters} between 2009 and 2011, and an associate editor for the \textsc{IEEE Transactions on Signal Processing} between 2012 and 2016. He has been a member of the IEEE Signal Processing for Communications and Networking Technical Committee (SPCOM-TC) since 2013, where he serves as chair 2019-2020. Since 2016 he is also responsible for the five year B.Sc and M.Sc.\ Degree Program in Electrical Engineering at KTH.

    For his work on MIMO communications, Joakim has been awarded the IEEE Signal Processing Society’s 2006 Young Author Best Paper Award, and the best student conference paper award at IEEE ICASSP 2007. He was also a recipient of the Ingvar Carlsson Career Award issued in 2009 by the Swedish Foundation for Strategic Research.  His recent work includes work on signal processing for biomedical data analysis, including single-cell tracking for time-lapse microscopy and inverse diffusion for immunoassays.
\end{IEEEbiography}

\enlargethispage{-.5cm}

\end{document}